\documentclass[11pt]{article}

\usepackage[english]{babel}
\usepackage{graphicx} % Figuren
\usepackage{amssymb,amsmath,amsfonts} % Heel belangrijk
\usepackage{stmaryrd} % Voor commando \llbracket
\usepackage{yhmath} % Voor commando \adots
\usepackage[latin1]{inputenc} % Om trema's en accenten te kunnen typen zonder speciaal commando
\usepackage[small,bf,hang]{caption} % Om caption bij tabellen en figuren in orde te krijgen
\usepackage{epic} % Voor de picture-omgeving

\def\beqa{\begingroup \setlength{\arraycolsep}{0.14em} \begin{eqnarray}}
\def\eeqa{\end{eqnarray} \endgroup \ignorespaces}
\def\beqas{\begingroup \setlength{\arraycolsep}{0.14em} \begin{eqnarray*}}
\def\eeqas{\end{eqnarray*} \endgroup \ignorespaces}

\newtheorem{theorem}{Theorem}%[chapter]

\graphicspath{{Figuren/}}

\begin{document}

\begin{center}
  {\Large \bf Angular momentum decomposition of the three-dimensional Wigner harmonic oscillator}\\[5mm]
  {\bf G.~Regniers \footnote{E-mail: Gilles.Regniers@UGent.be}, }
  {\bf and J.\ Van der Jeugt} \footnote{E-mail: Joris.VanderJeugt@UGent.be} \\[1mm]
  Department of Applied Mathematics and Computer Science, Ghent University, \\
  Krijgslaan 281-S9, B-9000 Gent, Belgium.
\end{center}

\vskip 10mm
\noindent
Short title: Wigner harmonic oscillator

\noindent
PACS numbers: 03.65.-w, 03.65.Fd, 02.20.Sv, 02.30.Gp

%----------
% ABSTRACT
%----------
\begin{abstract}
  \noindent In the Wigner framework, one abandons the assumption that the usual canonical commutation relations are necessarily valid. Instead, the compatibility of Hamilton's equations and the Heisenberg equations are the starting point, and no further assumptions are made about how the position and momentum operators commute. Wigner quantization leads to new classes of solutions, and representations of Lie superalgebras are needed to describe them. For the $n$-dimensional Wigner harmonic oscillator, solutions are known in terms of the Lie superalgebras $\mathfrak{osp}(1|2n)$ and $\mathfrak{gl}(1|n)$. For $n=3N$, the question arises as to how the angular momentum decomposition of representations of these Lie superalgebras is computed. We construct generating functions for the angular momentum decomposition of specific series of representations of $\mathfrak{osp}(1|6N)$ and $\mathfrak{gl}(1|3N)$, with $N=1$ and $N=2$. This problem can be completely solved for $N=1$. However, for $N=2$ only some classes of representations allow executable computations.
\end{abstract}

\newpage
\setcounter{equation}{0}

%--------------
% INTRODUCTION
%--------------
\section{Introduction}
The Hamiltonian of the $n$-dimensional non-isotropic harmonic oscillator with mass $m$ and frequencies $\omega_j \, (j=1, \ldots, n)$ is given by
\begin{equation} \label{ham-3Dwho} 
  \hat{H} = \frac{1}{2m} \sum_{j=1}^n \hat{p}_j^2 + \frac{m}{2} \sum_{j=1}^n \omega_j^2 \hat{q}_j^2,
\end{equation}
where the position and momentum operators are denoted by $\hat{p}_j$ and $\hat{q}_j$ respectively. We will look at this Hamiltonian in the framework of Wigner quantization, a concept which has been triggered by Wigner in~\cite{Wigner}, but was really introduced much later by Palev~\cite{Palev-79, Palev-82, Palev-86}. Since the Wigner perspective has already been considered in detail by Lievens and Van der Jeugt in~\cite{LVdJ-08}, we will only present their results succinctly. For a more thorough deduction of the results, we refer to the aforementioned paper. 

Wigner quantization dictates that the canonical commutation relations involving $\hat{p}_j$ and $\hat{q}_j$ should be replaced by less restrictive compatibility conditions (CCs). These arise by imposing the equivalence of Hamilton's equations and the Heisenberg equations. For the $n$-dimensional Wigner harmonic oscillator, the CCs are given by
\[
  [\hat{H}, \hat{q}_j] = - \frac{i \hbar}{m} \, \hat{p}_j, \qquad
  [\hat{H}, \hat{p}_j] = i \hbar m \omega_j^2 \, \hat{q}_j,
\]
for $j=1, \ldots, n$. By introducing new operators $a_j^\pm$ by
\begin{equation} \label{aj-pm}
  a_j^\pm = \sqrt{\frac{m \omega_j}{2 \hbar}} \hat{q}_j \mp \frac{i}{\sqrt{2m \hbar \omega_j}} \hat{p}_j
\end{equation}
we can rewrite the Hamiltonian as
\[
  \hat{H} = \frac{\hbar}{2} \sum_{j=1}^n \omega_j \{ a_j^+, a_j^- \}.
\]
In terms of these new operators, the compatibility conditions take the form
\begin{equation} \label{CCs-a-3Dwho}
  \sum_{j=1}^n \omega_j \bigl[ \{ a_j^+, a_j^- \}, a_k^\pm \bigr] = \pm 2 \omega_k a_k^\pm,
\end{equation}
for $k=1, \ldots, n$. Since the position and momentum operators are self-adjoint, we have $(a_j^\pm)^\dagger = a_j^\mp$. It turns out that we can find operators $a_j^\pm$ subject to the latter hermiticity conditions and to the compatibility conditions \eqref{CCs-a-3Dwho} in terms of Lie superalgebra generators. 

There are two known classes of solutions for the non-isotropic case; we can express $a_j^\pm$ in terms of elements of $\mathfrak{osp}(1|2n)$ and $\mathfrak{gl}(1|n)$. For each of these solutions, the spectrum of the Hamiltonian in specific Lie superalgebra representations was found in~\cite{LVdJ-08}. In their paper the authors give a nice overview of the relevant representations and their characters, and they present the energy spectrum by means of spectrum generating functions. For a detailed analysis, we refer the reader to that paper and the references therein. In this text we will summarize the elements of this paper that are useful for our purposes. It should be noted that we will restrict ourselves to the isotropic case where $\omega_j = \omega$ for all $j=1, \ldots, n$.

When both Lie superalgebra solutions are examined, it is time to move forward to the main objective of this paper. The solution of the Wigner quantum system under consideration depends on the Lie superalgebra representation $V$, in which the operators $a_j^\pm$ act. So the purpose is to study properties of the Wigner oscillator in different representations, one of which will correspond to the canonical case. For the three-dimensional $N$-particle Wigner harmonic oscillator, i.e. the case $n=3N$, we want to find the angular momentum/energy contents of Lie superalgebra representations of $\mathfrak{osp}(1|6N)$ and $\mathfrak{gl}(1|3N)$. In particular, this means we will try to discover which representations of $\mathfrak{so}(3)$ occur in the decomposition of specific Lie superalgebra representations at a given energy level. We will use the tool of generating functions to achieve this aim. The aspired results are known in the canonical case, and we will compare this case to the new solutions offered by Wigner quantization.

First, we discuss the orthosymplectic case in sections \ref{sec-osp-sol}, \ref{sec-ang-osp} and \ref{sec-gf-osp}. The $\mathfrak{osp}(1|2n)$ solution to the Wigner problem is discussed in section \ref{sec-osp-sol}. All of the results in this introductory section stem from~\cite{LVdJ-08}. In section \ref{sec-ang-osp} we explain how the angular momentum contents of the $\mathfrak{osp}(1|6N)$ representations $V(\mathfrak{p})$ can be found by means of generating functions, both theoretically as practically. The actual generating functions for representations of $\mathfrak{osp}(1|6)$ and $\mathfrak{osp}(1|12)$ are computed in section \ref{sec-gf-osp}. We also plot the angular momentum/energy contents in so-called $(E,j)$-diagrams and compare with the canonical case in this section. The Lie superalgebra solution $\mathfrak{gl}(1|n)$ is handled in sections \ref{sec-gl1n-sol}, \ref{sec-ang-gl} and \ref{sec-gf-gl}. In the concluding section, we summarize our main results.

%------------------------
% THE osp(1|2n) SOLUTION
%------------------------
\section{The $\mathfrak{osp}(1|2n)$ solution} \label{sec-osp-sol}
The orthosymplectic Lie superalgebra $\mathfrak{osp}(1|2n)$ is generated by its odd elements $b_j^\pm$ $(j=1, \ldots, n)$. These paraboson operators are subject to the so-called defining triple relations given by~\cite{Ganchev}
\begin{equation} \label{paraboson_triple}
  \bigl[ \{ b_j^\xi, b_k^\eta \}, b_l^\epsilon \bigr] =   (\epsilon - \xi) \delta_{jl} b_k^\eta 
                                                        + (\epsilon - \eta) \delta_{kl} b_j^\xi.
\end{equation}
In these triple relations, $j,k$ and $l$ are elements from the set $\{1,2,\ldots,n\}$ and $\eta, \xi, \epsilon \in \{+,-\}$ (to be interpreted as $+1$ and $-1$ in the algebraic expressions $(\epsilon - \xi)$ and $(\epsilon - \eta)$). The even elements of $\mathfrak{osp}(1|2n)$ are formed by taking anti-commutators $\{b_j^\xi, b_k^\eta\}$. 

We can use the paraboson operators to find solutions for the Wigner quantization discussed earlier. Indeed, writing $a_j^\pm$ as
\begin{equation}
  a_j^- = b_j^-, \qquad 
  a_j^+ = b_j^+,
\end{equation}
with $(j=1,2,\ldots,n)$, we see that the compatibility conditions \eqref{CCs-a-3Dwho} are satisfied using the defining triple relations \eqref{paraboson_triple}. The Hamiltonian \eqref{ham-3Dwho} then takes the following form:
\[
  \hat{H} = \frac{\hbar \omega}{2} \sum_{j=1}^n \{b_j^+, b_j^-\}.
\]
In order to obtain that $(a_j^\pm)^\dagger = a_j^\mp$, we need to work with suitable representations of $\mathfrak{osp}(1|2n)$. In the paraboson Fock space $V(\mathfrak{p})$ we automatically have $(b_j^\pm)^\dagger = b_j^\mp$, which makes this unitary irreducible representation of $\mathfrak{osp}(1|2n)$ an appropriate choice. In~\cite{LSVdJ-08-2} the representation $V(\mathfrak{p})$ was thoroughly investigated, resulting in an explicit basis, matrix elements and character formulas. The main theorem of that paper gives the conditions on $\mathfrak{p}$ for $V(\mathfrak{p})$ to be a unitary irreducible representation and it states the character of the representation~\cite[Theorem 7]{LSVdJ-08-2}.
\begin{theorem}
  The $\mathfrak{osp}(1|2n)$ representation $V(\mathfrak{p})$ with lowest weight $(\frac{\mathfrak{p}}{2}, \ldots, \frac{\mathfrak{p}}{2})$ is a unirrep if and only if $\mathfrak{p} \in \{ 1, \ldots, n-1 \}$ or $\mathfrak{p}>n-1$. \\
The character of $V(\mathfrak{p})$ is given by
\begin{equation} \label{char-osp-Vp}
  \mathrm{char} V(\mathfrak{p}) = (x_1 \cdots x_n)^{\mathfrak{p}/2} \sum_{\lambda, \, \ell(\lambda) \leq \lceil \mathfrak{p} \rceil} s_{\lambda}(x)
\end{equation}
The ceiling function $\lceil \mathfrak{p} \rceil$ is there to cover the cases where $n-1<\mathfrak{p}<n$.
\end{theorem}
We have used the notation $s_\lambda(x) = s_\lambda(x_1, \ldots, x_n)$ for the symmetric Schur function, which vanishes when the length of the partition $\lambda$, denoted by $\ell(\lambda)$, exceeds the number of variables $n$. The length of a partition is its number of parts. For a deep introduction to partitions and symmetric polynomials, we refer to Macdonald~\cite{MacDonald}.

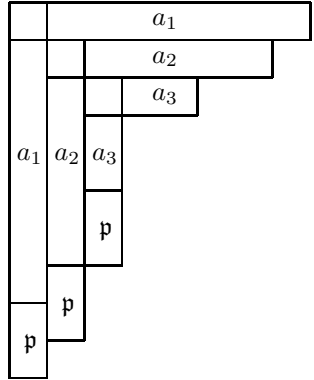
\begin{figure}[ht]
\vspace{0.5cm}
\setlength{\unitlength}{1mm}
\begin{center}
  \begin{picture}(1,1)
   % De lijnen
    \drawline(0,0)(0,-50)(5,-50)(5,-5)(10,-5)(10,-35)(15,-35)(15,-10)(35,-10)(35,-5)(10,-5)
    \drawline(0,0)(5,0)(5,-5)(0,-5)
    \drawline(5,0)(40,0)(40,-5)(35,-5)
    \drawline(5,-45)(10,-45)(10,-35)(5,-35)
    \drawline(10,-15)(25,-15)(25,-10)
    \drawline(5,-10)(15,-10)
    \drawline(10,-25)(15,-25)
    \drawline(0,-40)(5,-40)
   % De letters
    \put(19,-3){\small{$a_1$}}
    \put(19,-8){\small{$a_2$}}
    \put(19,-13){\small{$a_3$}}
    \put(1,-21){\small{$a_1$}}
    \put(6,-21){\small{$a_2$}}
    \put(11,-21){\small{$a_3$}}
    \put(12,-31){\small{$\mathfrak{p}$}}
    \put(7,-41){\small{$\mathfrak{p}$}}
    \put(2,-46){\small{$\mathfrak{p}$}}
  \end{picture}
\end{center}
\vspace{5.2cm}
\caption{Young diagram of the partition $\eta$ in Frobenius notation for $r=3$.}
\label{fig-Frob}
\end{figure}

For our purposes, the character formula \eqref{char-osp-Vp} is inadequate. Instead the following equivalent formula~\cite{LSVdJ-08-2} for $\mathfrak{p} \in \{ 1, 2, \ldots, n-1 \}$ will be more practical:
\begin{equation} \label{char-osp-E}
  \mathrm{char} V(\mathfrak{p}) = (x_1 \cdots x_n)^{\mathfrak{p}/2}
                                  \frac{\mathbf{E}_{(0, \mathfrak{p})}}{\prod_i (1-x_i) \prod_{j<k} (1-x_j x_k)},
\end{equation}
with
\[
  \mathbf{E}_{(0, \mathfrak{p})} = \sum_{\eta} (-1)^{c_\eta} s_\eta(x_1, \ldots, x_n).
\]
In this expression for $\mathbf{E}_{(0, \mathfrak{p})}$, the sum is over all partitions of the form
\[
  \eta = \begin{pmatrix}
                   a_1        &         a_2        & \cdots &         a_r        \\
           a_1 + \mathfrak{p} & a_2 + \mathfrak{p} & \cdots & a_r + \mathfrak{p}
         \end{pmatrix}
\]
in Frobenius notation, and 
\[
  c_\eta = a_1 + a_2 + \cdots + a_r + r.
\]
In the Frobenius notation for partitions, the first and second row denote the lengths of the rows and columns in the Young diagram of the partition, counted from the diagonal. For the partition $\eta$, a typical shape of the Young diagram is given in Figure \ref{fig-Frob}.

A simple expression for $\mathbf{E}_{(0, \mathfrak{p})}$ exists when $\mathfrak{p}=1$ and $\mathfrak{p}=n-1$. In these cases we have
\begin{equation} \label{char-V1}
  \mathrm{char} V(1) = (x_1 \cdots x_n)^{1/2} \frac{1}{\prod_i (1-x_i)}
\end{equation}
and
\begin{equation} \label{char-Vn-1}
  \mathrm{char} V(n-1) = (x_1 \cdots x_n)^{(n-1)/2}
                         \frac{(1-x_1 \cdots x_n)}{\prod_i (1-x_i) \prod_{j<k} (1-x_j x_k)}.
\end{equation}

It is possible to find the spectrum of the Hamiltonian $\hat{H}$ in the representation $V(\mathfrak{p})$. In fact, a straightforward technique described in~\cite{LVdJ-08} delivers the spectrum generating function $\mathrm{spec} \, \hat{H}$, which assigns all eigenvalues of $\hat{H}$ to a power of $t$. The multiplicity of an energy level $E$ is then given by the coefficient of $t^E$. For the $\mathfrak{osp}(1|2n)$ solution, the spectrum generating function takes the form
\begin{equation} \label{specH-osp}
  \mathrm{spec} \, \hat{H} = \sum_{k \geq 0} \;
                             \sum_{\lambda, \, |\lambda|=k, \, \ell(\lambda) \leq \lceil \mathfrak{p} \rceil} 
                             s_\lambda(1, \ldots, 1) \, t^{\hbar \omega (\frac{n \mathfrak{p}}{2} + k)},
\end{equation}
where $|\lambda|$ is the order of the partition, by which we mean the sum of its parts. The energy levels are equidistant and can be written as
\[
  E_k^{(\mathfrak{p})} = \hbar \omega (\frac{n \mathfrak{p}}{2} + k),
\]
with $ k=0,1,2, \ldots$. The multiplicities of these energy levels are equal to
\[
  \mu(E_k^{(\mathfrak{p})}) = \sum_{\lambda, \, |\lambda|=k, \, \ell(\lambda) \leq \lceil \mathfrak{p} \rceil}
                              s_\lambda(1, \ldots, 1).
%                           = \sum_{\lambda, \, |\lambda|=k, \, \ell(\lambda) \leq \lceil \mathfrak{p} \rceil}
%                             \binom{n}{\lambda'}.
\]
For $\mathfrak{p}=1$ all of the results above coincide with the canonical results. Indeed, $\mathfrak{p}=1$ represents canonical quantization.

%---------------------------------------------
% ANGULAR MOMENTUM DECOMPOSITION OF osp(1|2n)
%---------------------------------------------
\section{Angular momentum decomposition of $\mathfrak{osp}(1|2n)$} \label{sec-ang-osp}
The main objective of the present paper is to find the angular momentum content of Lie superalgebra representations related to the Wigner quantization of the $3D$ Wigner harmonic oscillator, both for $\mathfrak{osp}(1|2n)$ and $\mathfrak{gl}(1|n)$ with $n=3N$. Both cases are dissimilar with respect to the dimension of the representation spaces, so a proper approach is needed to tackle both problems. This asks for a small clarification.

We would like to describe the angular momentum content with a generating function. The representation $V(\mathfrak{p})$ of $\mathfrak{osp}(1|2n)$ is infinite-dimensional, which implies that it is impossible to construct a generating function comprising all $\mathfrak{osp}(1|2n)$ representations. Therefore, our objective in the $\mathfrak{osp}(1|2n)$ case is to construct a generating function for every representation $V(\mathfrak{p})$ separately. In the $\mathfrak{gl}(1|n)$ solution, examined from section \ref{sec-gl1n-sol} onwards, the representations will be finite-dimensional. In that case, the generating function will contain variables characterizing the $\mathfrak{gl}(1|n)$ representation.

We would like to find how the Hilbert space in which the Hamiltonian acts decomposes to $\mathfrak{so}(3)$ representations. In this section we will discuss the $\mathfrak{osp}(1|2n)$ case, where $n=3N$. In that case we can rely on the embedding
\begin{equation} \label{branching-osp1-6N-so3}
  \mathfrak{osp}(1|6N) \supset \mathfrak{sp}(6N)
             \supset \mathfrak{u}(3N)
             \supset \mathfrak{u}(3) \oplus \mathfrak{u}(N) 
             \supset \mathfrak{so}(3) \oplus \mathfrak{u}(1)
\end{equation}
to come up with a generating function that represents the angular momentum decomposition of the $\mathfrak{osp}(1|6N)$ representation $V(\mathfrak{p})$. Some explanation is needed to see why this is the correct embedding to use, i.e. to see why the angular momentum operators are elements of this $\mathfrak{so}(3)$ subalgebra of $\mathfrak{osp}(1|6N)$.

% Angular momentum
%------------------
\subsection{Angular momentum}
For $n=3$, our physical system is a three-dimensional harmonic oscillator. In a canonical context, the angular momentum operators are defined by $\mathbf{M} = \mathbf{\hat{q}} \times \mathbf{\hat{p}}$, or
\[
  M_j = \sum_{k,l=1}^3 \epsilon_{jkl} \, \hat{q}_k \hat{p}_l, \qquad (j=1,2,3),
\]
where $\epsilon_{jkl}$ is the Levi-Civita symbol. Since the position and momentum operator cannot be assumed to commute in Wigner quantization, a logical definition of the angular momentum operators is
\[
  M_j = \frac{1}{2} \sum_{k,l=1}^3 \epsilon_{jkl} \{\hat{q}_k, \hat{p}_l \}, \qquad (j=1,2,3),
\]
which, by means of \eqref{aj-pm}, can be written as
\begin{equation} \label{ang-mom-op-3}
  M_j = \frac{-i \hbar}{2} \sum_{k,l=1}^3 \epsilon_{jkl} \{a_k^+, a_l^- \}, \qquad (j=1,2,3).
\end{equation}
The compatibility conditions \eqref{CCs-a-3Dwho} do not contain enough information to lead to commutation relations between $M_1$, $M_2$ and $M_3$. However, in the $\mathfrak{osp}(1|6)$ solution $a_j^\pm = b_j^\pm$ one finds
\[
  [M_i, M_j] = i \hbar \, \epsilon_{ijk} M_k, \qquad (i,j,k = 1,2,3).
\]
These are the commutation relations of $\mathfrak{so}(3)$. Now, since the operators $M_j$ are in essence elements $\{b_k^+, b_l^- \}$ of $\mathfrak{osp}(1|6)$, they belong to the $\mathfrak{osp}(1|6)$ subalgebra $\mathfrak{u}(3)$ from the embedding
\[
  \mathfrak{osp}(1|6) \supset \mathfrak{sp}(6)
            \supset \mathfrak{u}(3)
            \supset \mathfrak{so}(3) \oplus \mathfrak{u}(1),
\]
which follows from~\cite[Proposition 3]{LSVdJ-08-2}. The generalization to $\mathfrak{osp}(1|6N)$ is rather straightforward. The physical system is now an $N$-particle three-dimensional harmonic oscillator. The position and momentum operators have a second index $\alpha$, with $\alpha=1, \ldots, N$. The angular momentum operators of the particle $\alpha$ can be written as
\[
  M_{j, \alpha} = \frac{1}{2} \sum_{k,l=1}^3 \epsilon_{jkl} \{\hat{q}_{k, \alpha}, \hat{p}_{l, \alpha} \}
                = \frac{-i \hbar}{2} \sum_{k,l=1}^3 \epsilon_{jkl} \{a_{k, \alpha}^+, a_{l, \alpha}^- \},
  \qquad (j=1,2,3).
\]
The total angular momentum is obtained by adding all the angular momenta of the individual particles. Thus we have
\begin{equation} \label{ang-mom-op-3N}
  M_j = \sum_{\alpha=1}^N  M_{j, \alpha}, \qquad (j=1,2,3).
\end{equation}
These $M_j$ are elements of the $\mathfrak{u}(3)$ subalgebra of $\mathfrak{osp}(1|6N)$ and satisfy the $\mathfrak{so}(3)$ commutation relations. Therefore the angular momentum components generate the $\mathfrak{so}(3)$ subalgebra of $\mathfrak{u}(3)$ in the following chain of subalgebras:
\[
  \mathfrak{osp}(1|6N) \supset \mathfrak{sp}(6N)
             \supset \mathfrak{u}(3N)
             \supset \mathfrak{u}(3) \oplus \mathfrak{u}(N) 
             \supset \mathfrak{so}(3) \oplus \mathfrak{u}(1).
\]
The question now is how the $\mathfrak{osp}(1|6N)$ representation $V(\mathfrak{p})$ decomposes with respect to these subalgebras.

% Decomposing the osp(1|6N) representation V(p)
%-----------------------------------------------
\subsection{Decomposing the $\mathfrak{osp}(1|6N)$ representation $V(\mathfrak{p})$} \label{subsec-decomp}
The starting point of the decomposition of $V(\mathfrak{p})$ is the character of the Lie superalgebra representation given by equation \eqref{char-osp-Vp}. Each Schur-function $s_\lambda(x_1,\ldots, x_n)$, with $\lambda = (\lambda_1, \ldots, \lambda_n)$ is the character of an irreducible covariant tensor representation of $\mathfrak{u}(n)$~\cite{Littlewood-40} and corresponds to the $\mathfrak{u}(n)$ representation with highest weight $\lambda$, where $n=3N$. In other words, equation \eqref{char-osp-Vp} is a $\mathfrak{u}(3N)$ character generating function. In contrast, we want the result of our analysis to be a representation generating function. In other words, the generating function returns all representations of $\mathfrak{so}(3)$ that appear in the decomposition of a fixed representation $V(\mathfrak{p})$ of $\mathfrak{osp}(1|6N)$. By means of an example we will try to avoid confusion between both concepts.

The character of $V(\mathfrak{p})$ given by equation \eqref{char-osp-Vp} is, as explained earlier, a $\mathfrak{u}(3N)$ character generating function. Indeed, it contains the characters of all $\mathfrak{u}(3N)$ representations in the decomposition of the $\mathfrak{osp}(1|6N)$ representation $V(\mathfrak{p})$. Such a $\mathfrak{u}(3N)$ character however, consist of many superfluous terms if one only wishes to know which $\mathfrak{u}(3N)$ representations appear. After all, a $\mathfrak{u}(3N)$ representation is characterized by a partition $\lambda$, so a term $x^\lambda = x_1^{\lambda_1} \ldots x_n^{\lambda_n}$ would suffice instead of $s_\lambda(x_1,\ldots, x_n)$.

Consider the representation $V(2)$ of $\mathfrak{osp}(1|6)$ for example. Following equation \eqref{char-Vn-1} we see that the $\mathfrak{u}(3)$ character generating function takes the form
\begin{equation} \label{char-gf-V2}
  \frac{x_1 x_2 x_3 (1-x_1 x_2 x_3)}{(1-x_1)(1-x_2)(1-x_3)(1-x_1 x_2)(1-x_1 x_3)(1-x_2 x_3)}.
\end{equation}
The expansion of this function contains all $\mathfrak{u}(3)$ characters in the decomposition of the $\mathfrak{osp}(1|6)$ representation $V(2)$. The $\mathfrak{u}(3)$ character generating function could just as well have been derived directly from equation \eqref{char-osp-Vp}. The partitions $\lambda$ in this equation have a maximum of two parts, so the $\mathfrak{u}(3)$ representation generating function is created by replacing every Schur function in equation \eqref{char-osp-Vp} by its leading term $x_1^{\lambda_1} x_2^{\lambda_2}$. We obtain
\begin{equation} \label{rep-gf-V2}
    x_1 x_2 x_3 \sum_{\lambda_2 = 0}^\infty \sum_{\lambda_1 = \lambda_2}^\infty x_1^{\lambda_1} x_2^{\lambda_2}
  = \frac{x_1 x_2 x_3}{(1-x_1)(1-x_1 x_2)}.
\end{equation}
Every monomial $x_1^{\lambda_1} x_2^{\lambda_2}$ in the expansion of this easier looking function corresponds to a $\mathfrak{u}(3)$ representation characterized by the partition $\lambda$. One can verify that the method described in the next paragraph, applied to the $\mathfrak{u}(3)$ character generating function \eqref{char-gf-V2}, will indeed give \eqref{rep-gf-V2} as a result.

Let us now return to the general case and consider the branching to $\mathfrak{u}(3) \oplus \mathfrak{u}(N)$ in \eqref{branching-osp1-6N-so3}. The substitution
\begin{equation}  \label{subs-gl3N-gl3glN}
  x_i := u_j v_l z, \qquad (j=1,2,3 \, \mbox{ and } \, l=1, \ldots, N)
\end{equation}
in equation \eqref{char-osp-E} yields a character generating function for $\mathfrak{u}(3) \oplus \mathfrak{u}(N)$. The factor
\[
  z=t^{\hbar \omega}
\]
keeps track of the energy, since the power of $z$ after the substitution \eqref{subs-gl3N-gl3glN} in \eqref{char-osp-Vp}  equals $|\lambda|$ and the order of the partition $\lambda$ determines the energy level $E_k^{(\mathfrak{p})}$, as can be seen from equation \eqref{specH-osp}. We will keep using this notation throughout the rest of the paper.

By now it should be clear that the $\mathfrak{u}(3N)$ character generating function \eqref{char-osp-E} is not a representation generating function. Likewise, after the substitution \eqref{subs-gl3N-gl3glN} one does not obtain a $\mathfrak{u}(3) \oplus \mathfrak{u}(N)$ representation generating function. Therefore, we need to describe a technique for changing a character generating function into a representation generating function.

% From character generating function to representation generating function
%- - - - - - - - - - - - - - - - - - - - - - - - - - - - - - - - - - - - - -
\subsubsection*{From character generating function to representation generating function}
A character generating function for a simple Lie algebra $\mathfrak{g}$ can generally be written as:
\[
  F(\eta) = \sum_\lambda \chi_\lambda(\eta) N_\lambda,
\]
where the sum runs over a fixed set of integrable highest weights of $\mathfrak{g}$, and where each character $\chi_\lambda(\eta)$ is the coefficient of a variable $N_\lambda$ of some sort. Suppose the vector $\eta = (\eta_1, \ldots, \eta_m)$ has $m$ components, corresponding to the number of nonzero Dynkin labels of $\lambda$. The Weyl character formula allows us to write the characters as
\[
  \chi_\lambda(\eta) = \frac{ \displaystyle{\sum_{w \in W}} \epsilon(w) \, \eta^{w(\lambda + \rho)} }
                            { \eta^\rho \displaystyle{\prod_{\alpha \in \Delta_+}} (1-\eta^{- \alpha}) },
\]
where $W$ is the Weyl group, $\rho$ is the Weyl tool and $\Delta_+$ is the set of positive roots of the Lie algebra. In order to transform $F(\eta)$ into a representation generating function, one has to multiply $F(\eta)$ by $\prod_{\alpha \in \Delta_+} (1-\eta^{- \alpha})$ and keep the terms in the dominant Weyl sector. All of this applied to our situation, where the Lie algebra $\mathfrak{u}(n)$ has the symmetric group as its Weyl group, means that we need to maintain the terms in $\eta^\lambda$, where $\lambda$ is a partition. One method of doing this, is by making the substitution
\[
  \eta_1 = c_1 \eta_1, \quad \eta_2 = \frac{c_2}{c_1} \, \eta_2, \quad \ldots \quad
  \eta_{m-1} = \frac{c_{m-1}}{c_{m-2}} \, \eta_{m-1}, \quad \eta_m = \frac{1}{c_{m-1}} \, \eta_m
\]
in $F(\eta) \prod_{\alpha \in \Delta_+} (1-\eta^{- \alpha})$ and keep all positive powers of $c_1, \ldots, c_m$. This comes down to finding the term in $c_1^0 \, c_2^0 \ldots c_{m-1}^0$ in the power series expansion of the function
\[
  F(\eta) \prod_{\alpha \in \Delta_+} (1-\eta^{- \alpha}) \frac{1}{ (1-c_1^{-1}) \cdots (1-c_m^{-1}) }.
\]
Several computational software packages have specific methods of finding constant terms in an expression.

For our $\mathfrak{u}(3) \oplus \mathfrak{u}(N)$ character generating function, we would have to perform the substitution described above for the variables $u_i$, belonging to $\mathfrak{u}(3)$, and $v_l$, belonging to $\mathfrak{u}(N)$. Since the next step in the decomposition \eqref{branching-osp1-6N-so3} is from $\mathfrak{u}(3) \oplus \mathfrak{u}(N)$ to $\mathfrak{so}(3) \oplus \mathfrak{u}(1)$, we want the $\mathfrak{u}(N)$ representation labels to be replaced by the dimension of the corresponding $\mathfrak{u}(N)$ representation in the obtained representation generating function for $\mathfrak{u}(3) \oplus \mathfrak{u}(N)$.

% Introducing the dimensions of the u(N) representations
%- - - - - - - - - - - - - - - - - - - - - - - - - - - - -
\subsubsection*{Introducing the dimensions of the $\mathfrak{u}(N)$ representations}
Replacing a term $v^\nu = v_1^{\nu_1} \ldots v_N^{\nu_N}$ by the dimension of the $\mathfrak{u}(N)$ representation labeled by $\nu = (\nu_1, \ldots, \nu_N)$ demands knowledge of a so-called dimension generating function, in which the coefficient of $v^\nu$ is the dimension of the corresponding $\mathfrak{u}(N)$ representation. The dimension of such a representation is known~\cite{MacDonald} and equals $s_\nu(1, \ldots, 1)$. Thus, the dimension generating function we need is of the form
\[
  \sum_\nu s_\nu(1, \ldots, 1) v^\nu.
\]
An expression for this $\mathfrak{u}(N)$ dimension generating function for general $N$ is not known. However, for our purposes the $\mathfrak{u}(2)$ dimension generating function will be enough. In this $N=2$ case we have
\begin{equation} \label{dim-gf-u2-x1x2}
  \sum_\nu s_\nu(x_1, x_2) v^\nu = \frac{1}{(1-x_1 v_1) (1-x_2 v_1) (1-x_1 x_2 v_1 v_2)}.
\end{equation}
Indeed, the summation on the left-hand side can be written as
\[
  \sum_{\lambda_1 = \lambda_2}^\infty \sum_{\lambda_2 = 0}^\infty \sum_{k=0}^{\lambda_1 - \lambda_2}
  x_1^{\lambda_1 -k} x_2^{\lambda_2 +k} v_1^{\lambda_1} v_2^{\lambda_2},
\]
which simplifies to the right-hand side of \eqref{dim-gf-u2-x1x2}. Therefore, the dimension generating function of $\mathfrak{u}(2)$ is
\begin{equation} \label{dim-gf-u2}
  \frac{1}{(1-v_1)^2 (1-v_1 v_2)}.
\end{equation}
We now want this $\mathfrak{u}(2)$ dimension generating function and the previously obtained $\mathfrak{u}(3) \oplus \mathfrak{u}(2)$ generating function -- let us denote this by $H_2(u,v)$ -- to be ``substituted'' in one another. Saying that two generating functions $F_1(X,Y)$ and $F_2(X,Z)$, with common variables $X$, are substituted in each other, means that $X^x$ is replaced in either of these generating functions by the coefficient of $X^x$ in the other. This is achieved by finding the term in $X^0$ in either $F_1(X,Y) F_2(X^{(-1)},Z)$ or $F_1(X^{(-1)},Y) F_2(X,Z)$, whichever is more easily calculated. Indeed, if $F_1(X,Y)$ contains a term $p_1(Y) X^x$ and $F_2(X^{(-1)},Z)$ includes a term $p_2(Z) X^{(-x)}$, then we find a term $p_1(Y) p_2(Z)$ in the product of both functions.

Here, we have $H_2(u,v)$ on the one hand and the $\mathfrak{u}(2)$ dimension generating function \eqref{dim-gf-u2} on the other hand. Substituting these generating functions in each other is done by taking the constant term in the variables $v_1$ and $v_2$ in the expansion of
\[
  H_2(u,v) \frac{1}{(1-v_1^{-1})^2 (1-v_1^{-1} v_2^{-1})}.
\]
This replaces the variables $v_1$ and $v_2$ in $H_2(u,v)$ by the dimensions of the corresponding $\mathfrak{u}(2)$ representations. Again, finding the constant term can be done by various mathematical software packages.

% Angular momentum content
%- - - - - - - - - - - - - -
\subsubsection*{Angular momentum content}
What is left now is a generating function which models the decomposition of a $\mathfrak{u}(3N)$ representation into $\mathfrak{u}(3)$ representations. These representations of $\mathfrak{u}(3)$ are labeled by the variables $u_1, u_2, u_3$, the powers of which represent the partition $(\lambda_1, \lambda_2, \lambda_3)$ that characterizes the representation. Further decomposition to $\mathfrak{so}(3)$ is brought about by the known generating function for $\mathfrak{su}(3) \supset \mathfrak{so}(3)$~\cite{Gaskell-78}, given here in Dynkin label notation:
\begin{equation} \label{genf-sl3-so3-dl}
  \frac{1+PQJ}{(1-PJ)(1-QJ)(1-P^2)(1-Q^2)},
\end{equation}
in which $J$ is the $\mathfrak{so}(3)$ label. Hence, the generating function for $\mathfrak{u}(3) \supset \mathfrak{so}(3)$ in partition notation can be written as
\begin{equation} \label{genf-sl3-so3-pl}
  G(u_1, u_2, u_3) = \frac{1 + u_1^2 u_2 \, J}
                          {(1 - u_1 u_2 u_3)(1 - u_1 \, J)(1 - u_1 u_2 \, J)
                           (1 - u_1^2)(1 - u_1^2 u_2^2)}.
\end{equation}
All factors from equation \eqref{genf-sl3-so3-dl} appear in \eqref{genf-sl3-so3-pl} in accordance with the relation $[p,q] = [\lambda_1 - \lambda_2, \lambda_2 - \lambda_3]$, except for $(1 - u_1 u_2 u_3)$ in the denominator of \eqref{genf-sl3-so3-pl}. This is explained by the fact that the Dynkin-label $[p,q]$ is not influenced when a random integer is added to every part of the partition $\lambda$. 

Substituting one of these generating functions into the other is done by a similar technique as before. One just has to multiply the first generating function, embodying the embedding of $\mathfrak{u}(3)$ in $\mathfrak{osp}(1|6N)$, by $G(u_1^{-1}, u_2^{-1}, u_3^{-1})$ and take the term in $u_1^0 u_2^0 u_3^0$ in this expression. The resulting generating function describes the angular momentum content of the $\mathfrak{osp}(1|6N)$ representation $V(\mathfrak{p})$.

%-------------------------------------------------
% GENERATING FUNCTIONS FOR osp(1|6) AND osp(1|12)
%-------------------------------------------------
\section{Generating functions for $\mathfrak{osp}(1|6)$ and $\mathfrak{osp}(1|12)$} \label{sec-gf-osp}
Remember that the goal in the orthosymplectic case is to derive a generating function for each representation separately. Such a generating function will be a function of two variables, $J$ and $z$. The former labels the $\mathfrak{so}(3)$ content of the representation $V(\mathfrak{p})$, while the latter accounts for the $\mathfrak{u}(1)$ part. In fact, each power of $z$ stands for an energy level.

We can now apply the techniques described in the previous section to derive generating functions for the angular momentum decomposition of the representations $V(\mathfrak{p})$ of $\mathfrak{osp}(1|6)$ and $\mathfrak{osp}(1|12)$. However, the $\mathfrak{osp}(1|6)$ case simplifies drastically as each representation $V(\mathfrak{p})$ decomposes to $\mathfrak{u}(3)$ right away. Therefore we can use a different logic to find the desired generating functions.

% Generating functions for osp(1|6) > so(3) + u(1)
%-------------------------------------------------
\subsection{Generating functions for $\mathfrak{osp}(1|6) \supset \mathfrak{so}(3) \oplus \mathfrak{u}(1)$}
From the character formula in equation \eqref{char-osp-Vp} one can see that the representation $V(\mathfrak{p})$ of $\mathfrak{osp}(1|6)$ decomposes as a direct sum of $\mathfrak{u}(3)$ representations labeled by a partition $\lambda$, where $\lambda$ has at most three parts. The branching of these $\mathfrak{u}(3)$ representations can immediately be obtained with the help of equation \eqref{genf-sl3-so3-pl}. We separate three cases: $\mathfrak{p}=1$, $\mathfrak{p}=2$ and $\mathfrak{p} > 2$.

% p=1
%- - -
$\mathbf{\mathfrak{p}=1: }$ 
All partitions in the character formula \eqref{char-osp-Vp} have length 1, so $\lambda_2 = \lambda_3 = 0$. It is then obvious that the generating function for $\mathfrak{p}=1$ is simply $G(z,0,0)$, where $G(u_1, u_2, u_3)$ is the generating function \eqref{char-osp-Vp}. Not forgetting the factor $z^{3 \mathfrak{p}/2}$ for the energy we obtain
\begin{equation} \label{genf-osp16-p1}
  \frac{z^{3/2}}{(1-zJ)(1-z^2)}.
\end{equation}
One can use this generating function to derive the $\mathfrak{so}(3)$ representations that emerge at energy level $E_k^{(1)}$. This information can be made accessible by means of a table in which the element in row $k+1$ and column $j+1$ (counted from the bottom) marks the number of representations $J^j$ at energy level $E_k^{(1)}$ in the angular momentum decomposition of $\mathfrak{osp}(1|6)$. We call this the $(E,j)$-diagram of $\mathfrak{osp}(1|6)$ for $\mathfrak{p}=1$.
\[
  \begin{tabular}{cc|cccccc}
    $\vdots$ &     &   &   &   &   &   & $\adots$ \\
      11/2   &     & 1 &   & 1 &   & 1 &          \\
       9/2   &     &   & 1 &   & 1 &   &          \\
       7/2   &     & 1 &   & 1 &   &   &          \\
       5/2   &     &   & 1 &   &   &   &          \\
       3/2   &     & 1 &   &   &   &   &          \\ \hline
      $E_k$  &     &   &   &   &   &   &          \\
             & $j$ & 0 & 1 & 2 & 3 & 4 & $\cdots$
  \end{tabular}
\]
Indeed, the first few terms in the expansion of \eqref{genf-osp16-p1} are
\[
  z^{3/2} + J \, z^{5/2} + (1+J^2) \, z^{7/2} + (J+J^3) \, z^{9/2} + (1+J^2+J^4) \, z^{11/2} + \cdots.
\]
We see for example that at energy level $E_k^{(1)} = 9/2 \hbar \omega$, there are two $\mathfrak{so}(3)$ representations in the decomposition of the $\mathfrak{osp}(1|6)$ representation $V(1)$, characterized by $j=1$ and $j=3$. Of course, these results were already known because $\mathfrak{p} = 1$ represents the canonical case. This $(E,j)$-diagram for instance also appears in~\cite{Wybourne-74}.

% p=2
%- - -
$\mathbf{\mathfrak{p}=2: }$ The partition $\lambda$ now has at most two parts, so $\lambda_3 = 0$. The $\mathfrak{so}(3) \oplus \mathfrak{u}(1)$ decomposition of $\mathfrak{u}(3)$ representations labeled by such partitions is given by $G(z,z,0)$. Therefore, we can write the generating function for the angular momentum decomposition of $\mathfrak{osp}(1|6)$ for $\mathfrak{p}=2$ as
\begin{equation} \label{genf-osp16-p2}
  \frac{(1+z^3J) \, z^3}{(1-zJ)(1-z^2J)(1-z^2)(1-z^4)}.
\end{equation}
As in the previous case, we can generate the $(E,j)$-diagram of $\mathfrak{osp}(1|6)$ for $\mathfrak{p}=2$.
\[
  \begin{tabular}{cc|cccccc}
    $\vdots$ &     &   &   &   &   &   & $\adots$ \\
        7    &     & 2 & 1 & 3 & 1 & 1 &          \\
        6    &     &   & 2 & 1 & 1 &   &          \\
        5    &     & 1 & 1 & 1 &   &   &          \\
        4    &     &   & 1 &   &   &   &          \\
        3    &     & 1 &   &   &   &   &          \\ \hline
      $E_k$  &     &   &   &   &   &   &          \\
             & $j$ & 0 & 1 & 2 & 3 & 4 & $\cdots$
  \end{tabular}
\]
Let us look at the case $E_k=7 \hbar \omega$, i.e. $k=4$ as an example. There are three partitions with two parts of order $4$, namely $(4,0,0)$, $(3,1,0)$, and $(2,2,0)$. The $\mathfrak{so}(3)$ representations that emerge in these cases can be found by equation \eqref{genf-sl3-so3-pl}. In total we have
\[
  (1+J^2+J^4) + (J+J^2+J^3) + (1+J^2).
\]
This is in accordance with the coefficient of $z^7$ in equation \eqref{genf-osp16-p2}, as can be seen from the $(E,j)$-diagram as well.

% p>2
%- - -
$\mathbf{\mathfrak{p} > 2: }$ Since the length of the partitions in \eqref{char-osp-Vp} cannot exceed the number of variables, we are looking in this case at partitions of length at most $3$. So the generating function for $\mathfrak{p} > 2$ is $z^{3 \mathfrak{p}/2} \, G(z,z,z)$, or
\begin{equation} \label{genf-osp16-p3}
  \frac{(1+z^3J) \, z^{3 \mathfrak{p}/2}}{(1-zJ)(1-z^2J)(1-z^2)(1-z^3)(1-z^4)}.
\end{equation}
The $(E,j)$-diagram for $\mathfrak{p} > 2$ is given by.
\[
  \begin{tabular}{cc|cccccc}
           $\vdots$       &     &   &   &   &   &   & $\adots$ \\
    $3\mathfrak{p}/2 + 4$ &     & 2 & 2 & 3 & 1 & 1 &          \\
    $3\mathfrak{p}/2 + 3$ &     & 1 & 2 & 1 & 1 &   &          \\
    $3\mathfrak{p}/2 + 2$ &     & 1 & 1 & 1 &   &   &          \\
    $3\mathfrak{p}/2 + 1$ &     &   & 1 &   &   &   &          \\
    $3\mathfrak{p}/2$     &     & 1 &   &   &   &   &          \\ \hline
             $E_k$        &     &   &   &   &   &   &          \\
                          & $j$ & 0 & 1 & 2 & 3 & 4 & $\cdots$
  \end{tabular}
\]
Notice that for the lower energy levels, the cases $\mathfrak{p} = 2$ and $\mathfrak{p} > 2$ do not differ very much from the canonical case. The larger discrepancies are found in higher energy regions.

% Generating functions for osp(1|12) > so(3) + u(1)
%--------------------------------------------------
\subsection{Generating functions for $\mathfrak{osp}(1|12) \supset \mathfrak{so}(3) \oplus \mathfrak{u}(1)$}
The previous case might have been very elementary, for $\mathfrak{osp}(1|12)$ the computations are much harder. In fact, it is not practically possible to find generating functions for all $\mathfrak{osp}(1|12)$ representation $V(\mathfrak{p})$. For $V(1)$ and $V(2)$ however, we are able to follow all the steps from section \ref{subsec-decomp} to construct the generating function for the angular momentum decomposition. We were unable to compute these generating functions for the representations $V(\mathfrak{p})$ with $\mathfrak{p} \geq 3$.

% The representation V(1)
%- - - - - - - - - - - - -
\subsubsection*{The representation $V(1)$}
We start with the character of $V(1)$, given by equation \eqref{char-V1}, and perform the substitution \eqref{subs-gl3N-gl3glN}, thus creating a $\mathfrak{u}(3) \oplus \mathfrak{u}(2)$ character generating function:
\[
  \frac{(u_1^2 u_2^2 u_3^2 \, v_1 v_2)^{3/2} z^3}
       {(1-u_1 v_1 z)(1-u_2 v_1 z)(1-u_3 v_1 z)(1-u_1 v_2 z)(1-u_2 v_2 z)(1-u_3 v_2 z)}.
\]
We need to change this into a representation generating function. To this end, we multiply the previous function by
\[
  \prod_{\alpha \in \Delta_+} (1-u^{- \alpha}) \prod_{\alpha' \in \Delta'_+} (1-v^{- \alpha'}),
\]
where $\Delta_+$ and $\Delta'_+$ are the positive roots of $\mathfrak{u}(3)$ and $\mathfrak{u}(2)$ respectively. Thus we have
\[
  \Delta_+ = \{ (1,-1,0), (1,0,-1), (0,1,-1) \} \qquad \mbox{ and } \qquad 
  \Delta'_+ = \{ (1,-1) \}.
\]
Therefore we need to multiply our $\mathfrak{u}(3) \oplus \mathfrak{u}(2)$ character generating function by
\[
  (1-\frac{u_2}{u_1})(1-\frac{u_3}{u_1})(1-\frac{u_3}{u_2}) (1-\frac{v_2}{v_1})
\]
and perform the substitutions 
\[
         u_1 = a u_1, \quad u_2 = \frac{b u_2}{a}, \quad u_3 = \frac{u_3}{b},
  \qquad v_1 = c v_1, \quad v_2 = \frac{v_2}{c}.
\]
We want to keep all positive powers of $a$, $b$ and $c$, so we multiply our function by
\[
  \frac{1}{(1-a^{-1})(1-b^{-1})(1-c^{-1})}
\]
and find the constant term in $a$, $b$ and $c$. This is the hardest step to compute. The term in $a^0 b^0 c^0$ factorizes nicely as
\[
  \frac{(u_1^2 u_2^2 u_3^2 \, v_1 v_2)^{3/2} z^3}
       {(1-u_1 v_1 z)(1-u_1 u_2 v_1 v_2 z^2)}.
\]
This is the $\mathfrak{u}(3) \oplus \mathfrak{u}(2)$ representation generating function, in which we want to change the $\mathfrak{u}(2)$ labels $v_1$ and $v_2$ by the dimensions of the corresponding $\mathfrak{u}(2)$ representations. Following equation \eqref{dim-gf-u2}, the $\mathfrak{u}(2)$ dimension generating function is
\[
  \frac{1}{(1-v_1)^2(1-v_1 v_2)}.
\]
Substituting this into the previously obtained $\mathfrak{u}(3) \oplus \mathfrak{u}(2)$ representation generating function gives
\[
  \frac{u_1 u_2 u_3 \, z^3}{(1-u_1 z)^2 (1-u_1 u_2 z^2)}.
\]
The angular momentum content is then found by substituting this function and $G(u_1, u_2, u_3)$ from equation \eqref{genf-sl3-so3-pl} into each other. The resulting angular momentum generating function for the representation $V(1)$ of $\mathfrak{osp}(1|12)$ is
\begin{equation}
  \frac{(1+Jz^2) \, z^3}{(1-z^2)^3 (1-Jz)^2}.
\end{equation}
The $(E,j)$-diagram shows the result for the first few energy levels.
\[
  \begin{tabular}{cc|cccccc}
    $\vdots$ &     &   &   &   &   &   & $\adots$ \\
      11/2   &     & 6 & 3 & 9 & 3 & 5 &          \\
       9/2   &     &   & 6 & 2 & 4 &   &          \\
       7/2   &     & 3 & 1 & 3 &   &   &          \\
       5/2   &     &   & 2 &   &   &   &          \\
       3/2   &     & 1 &   &   &   &   &          \\ \hline
      $E_k$  &     &   &   &   &   &   &          \\
             & $j$ & 0 & 1 & 2 & 3 & 4 & $\cdots$
  \end{tabular}
\]
%

% The representation V(2)
%- - - - - - - - - - - - -
\subsubsection*{The representation $V(2)$}
The character of $V(2)$, obtained from \eqref{char-osp-E}, does not factorize nicely. The angular momentum generating function for $V(2)$ can be constructed in precisely the same manner as for $V(1)$. The $\mathfrak{u}(3) \oplus \mathfrak{u}(2)$ representation generating function in this case equals $N/D$, with
\begin{subequations} \label{u3-u2-rep-gf-V2}
  \begin{equation}
    \begin{split}
      N = \, 1 & + \, u^{21}v^{21}z^3 + u^{221}v^{32}z^5 + u^{321}v^{42}z^6 
                                      + u^{321}v^{33}z^6 - 2u^{321}v^{42}z^6          \\
               & - \, u^{421}v^{52}z^7 - u^{421}v^{43}z^7 - u^{422}v^{53}z^8 
                                       - u^{431}v^{53}z^8 - u^{432}v^{63}z^9          \\
               & - \, u^{432}v^{54}z^{9} - 2u^{532}v^{64}z^{10} + u^{532}v^{73}z^{10} 
                                         + u^{532}v^{64}z^{10} + u^{632}v^{74}z^{11}  \\
               & + \, u^{643}v^{85}z^{13} + u^{853}v^{10,6}z^{16}
    \end{split}
  \end{equation}
  and
  \begin{equation}
    \begin{split}
      D = \;\, & (1 - u^{1}v^{1}z) (1 - u^{2}v^{11}z^2) (1 - u^{11}v^{2}z^2)
                                   (1 - u^{11}v^{11}z^2) (1 - u^{111}v^{21}z^3)                              \\
               & (1 - u^{22}v^{22}z^4) (1 - u^{211}v^{31}z^4) (1 - u^{211}v^{22}z^4) (1 - u^{222}v^{33}z^6),
    \end{split}
  \end{equation}
\end{subequations}
where we have used the notation $u^\lambda = u_1^{\lambda_1} u_2^{\lambda_2} u_3^{\lambda_3}$, and similarly for $v$. The powers of $z$ are integers, not partitions. The angular momentum generating function, also quite cumbersome, has a numerator equal to
\beqas
 z^6 \, \bigl( 1 & - \, & 2z + 3z^2+Jz^2 - 2z^3+4Jz^3 + 6z^4-7Jz^4-3J^2z^4-J^3z^4 - 6z^5+4Jz^5-2J^2z^5             \\
                 & - \, & 4J^3z^5 + 6z^6-8Jz^6-3J^2z^6+10J^3z^6+3J^4z^6 - 2z^7+10Jz^7+4J^2z^7-8J^3z^7              \\
                 & + \, & 3z^8-14Jz^8-16J^2z^8+13J^3z^8-3J^4z^8-3J^5z^8 - 2z^9+2Jz^9+6J^2z^9-12J^3z^9              \\
                 & + \, & 6J^4z^9+4J^5z^9 + z^{10}-5Jz^{10}+2J^2z^{10}+28J^3z^{10}+2J^4z^{10}-5J^5z^{10}+J^6z^{10} \\
                 & + \, &   4Jz^{11}+6J^2z^{11}-12J^3z^{11}+6J^4z^{11}+2J^5z^{11}-2J^6z^{11}
                          - 3Jz^{12}-3J^2z^{12}+13J^3z^{12}                                                        \\
                 & - \, &   16J^4z^{12}-14J^5z^{12}+3J^6z^{12} 
                          - 8J^3z^{13}+4J^4z^{13}+10J^5z^{13}-2J^6z^{13} + 3J^2z^{14}                              \\
                 & + \, &   10J^3z^{14}-3J^4z^{14}-8J^5z^{14}+6J^6z^{14} 
                          - 4J^3z^{15}-2J^4z^{15}+4J^5z^{15}-6J^6z^{15} - J^3z^{16}                                \\
                 & - \, &   3J^4z^{16}-7J^5z^{16}+6J^6z^{16} + 4J^5z^{17}-2J^6z^{17}
                       + J^5z^{18}+3J^6z^{18} - 2J^6z^{19} + J^6z^{20}            \bigr),
\eeqas
while the denominator is
\[
  (1-z^4)^4 (1-z^2)^2 (1-z)^2 (1-J^2z^2) (1-Jz^2)^3 (1-Jz)^2.
\]
Schematically thrown into a $(E,j)$-diagram this gives
\[
  \begin{tabular}{cc|cccccc}
    $\vdots$ &     &    &   &   &   &   & $\adots$ \\
        7    &     & 19 & 22 & 34 & 15 & 9 &          \\
        6    &     &  2 & 14 &  8 &  6 &   &          \\
        5    &     &  4 &  4 &  4 &    &   &          \\
        4    &     &    &  2 &    &    &   &          \\
        3    &     &  1 &    &    &    &   &          \\ \hline
      $E_k$  &     &    &    &    &    &   &          \\
             & $j$ &  0 &  1 &  2 &  3 & 4 & $\cdots$
  \end{tabular}
\]
for the lower energies.

%----------------------
% THE gl(1|n) SOLUTION
%----------------------
\section{The $\mathfrak{gl}(1|n)$ solution} \label{sec-gl1n-sol}
The elements of the general linear Lie superalgebra $\mathfrak{gl}(1|n)$ are denoted by $e_{jk}$, $j,k = 0,1, \ldots, n$, where $e_{0j}$ and $e{j0}$ are the odd generators. The commutation and anti-commutation relations valid for this algebra are
\begin{equation}
  \llbracket e_{ij}, e_{kl} \rrbracket = \delta_{jk} e_{il} - (-1)^{deg(e_{ij}) deg(e_{kl})} \delta_{il} e_{kj}.
\end{equation}
In terms of $\mathfrak{gl}(1|n)$ generators, solutions of the compatibility conditions \eqref{CCs-a-3Dwho} can be written as
\[
  a_j^- = \sqrt{\frac{2}{n-1}} \, e_{j0},                     \qquad
  a_j^+ = \sqrt{\frac{2}{n-1}} \, e_{0j}.
\]
The hermiticity condition $(a^\pm)^\dagger = a^\mp$ implies the star condition
\begin{equation} \label{star-cond-gl1n}
  (e_{0j})^\dagger = e_{j0}.
\end{equation}
The Hamiltonian can then be rewritten as
\[
  \hat{H} = \frac{\hbar \omega}{n-1} \bigl( n e_{00} + \sum_{j=1}^n e_{jj} \bigr).
\]
The unitary representations of $\mathfrak{gl}(1|n)$ compatible with the star condition \eqref{star-cond-gl1n} are known~\cite{Gould}: aside from the typical representations, we have the covariant and contravariant tensor representations. Here we are going to work with the covariant tensor representations $V_\lambda$, labeled by a partition $\lambda$ with $\lambda_2 \leq n$. The character of this representation was given by Berele and Regev in~\cite{Berele-87}. It is a supersymmetric Schur function $s_\lambda(x_1|y_1, \ldots, y_n)$ that can be written as
\beqa 
  \mathrm{char} V_\lambda & = & s_\lambda(x_1|y_1, \ldots, y_n)                         \nonumber \\
                          & = & \sum_{\mu, \nu} c_{\mu \nu}^\lambda 
                                                s_\mu (x_1) s_{\nu'}(y_1, \ldots, y_n), \label{char-gl-Vlambda}
\eeqa
where the coefficients $c_{\mu \nu}^\lambda$ are the Littlewood-Richardson coefficients and $\nu'$ is the conjugate partition of $\nu$, i.e. the partition that is obtained when the Young diagram of $\nu$ is transposed. The Littlewood-Richardson coefficients are integers and arise as coefficients in the expansion of a product of two Schur functions as a linear combination of Schur functions.

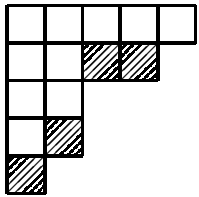
\begin{figure}[ht]
\vspace{0.5cm}
\setlength{\unitlength}{1mm}
\begin{center}
  \begin{picture}(1,1)
   % Gestreept blokje definitie
   % Er wordt getekend vanaf de linkerbovenhoek van het blokje
    \newsavebox{\streepblokje}
    \savebox{\streepblokje}(0, -5)[l]{
      \drawline(0,0)(5,0)(5,-5)(0,-5)(0,0)
      \multiput(0,0)(4,-4){2}{\drawline(1,0)(0,-1)}
      \multiput(0,0)(3,-3){2}{\drawline(2,0)(0,-2)}
      \multiput(0,0)(2,-2){2}{\drawline(3,0)(0,-3)}
      \multiput(0,0)(1,-1){2}{\drawline(4,0)(0,-4)}
      \drawline(5,0)(0,-5)     
    }
   % Plaatsen gestreepte blokjes
    \put(0,-20){\usebox{\streepblokje}}
    \put(5,-15){\usebox{\streepblokje}}
    \put(10,-5){\usebox{\streepblokje}}
    \put(15,-5){\usebox{\streepblokje}}
   % Partitie
    \thicklines
    \drawline(0,-20)(10,-20)(10,-10)(20,-10)(20,-5)(25,-5)(25,0)(0,0)(0,-25)(5,-25)(5,-15)(10,-15)
    \drawline(0,-15)(5,-15)(5,0)
    \drawline(0,-5)(20,-5)(20,0)
    \drawline(0,-10)(10,-10)(10,0)
    \drawline(15,0)(15,-10)
  \end{picture}
\end{center}
\vspace{2.2cm}
\caption{The horizontal $4$-strip $\lambda - \nu$ with $\lambda = (5,4,2,2,1)$ and $\nu = (5,2,2,1)$.}
\label{fig-hor-strip}
\end{figure}

Equation \eqref{char-gl-Vlambda} is worth a closer look. First, we note that $s_\mu (x_1)$ vanishes unless the length of the partition $\mu$ is equal to one, the number of variables of the Schur function. Thus, only partitions of the form $\mu = (r)$ are allowed. In this case we have
\[
  s_{(r)}(x_1) = x_1^r.
\]
The Littlewood-Richardson coefficients simplify a lot in this case as well. In~\cite[\textsection 5]{MacDonald} we find
\[
  c_{(r) \nu}^\lambda = \begin{cases}
                          1 & \mbox{if $\lambda - \nu$ is a horizontal $r$-strip} \\
                          0 & \mbox{otherwise}.
                        \end{cases}
\]
To explain what a horizontal $r$-strip is, we must first introduce the notion of a skew diagram. Consider two partitions $\lambda$ and $\nu$ such that $\nu_j \leq \lambda_j$ for all $j$. In other words, the Young diagram of $\nu$ is embedded in the Young diagram of $\lambda$. The set-theoretic difference $\theta = \lambda - \nu$ is called a skew diagram and contains the squares that belong to the Young diagram of $\lambda$ but not to the Young diagram of $\nu$. If $\theta$ contains at most one block per column, i.e. $\theta_i' \leq 1$, the skew diagram is called a horizontal strip. A horizontal strip with $r$ blocks is then called a horizontal $r$-strip. In Figure \ref{fig-hor-strip} we find an example of a horizontal $4$-strip.

Combining these results, we find that equation \eqref{char-gl-Vlambda} can conveniently be rewritten as
\[
  \mathrm{char} V_\lambda = s_\lambda(x_1| \, y) = \sum_{r \geq 0} x_1^r \sum_\nu s_{\nu'}(y),
\]
where the second summation runs over all partitions $\nu$ such that $\lambda - \nu$ is a horizontal $r$-strip. The first summation is not infinite. Since the Young diagram of the partition $\lambda$ has $\lambda_1$ columns, the horizontal strip $\lambda - \nu$ can only have a maximum of $\lambda_1$ parts. So $r$ will never exceed the value $\lambda_1$. Another remark is that $r$ cannot be too small either. If $\lambda - \nu$ would be an $r$-strip with $r < \lambda_1 - n$, then $\nu$ would have to be a partition with $\nu_1 > n$. This would imply that the length of the conjugate partition $\nu'$ is larger than $n$, which means that the Schur function $s_{\nu'}(y)$ vanishes. Since $\lambda_1 - n$ could be negative, we say that $r$ takes values between $r^*$ and $\lambda_1$, where $r^*$ is given by 
\[
  r^* = \begin{cases}
                0       & \mbox{if $\lambda_1 \leq n$} \\
          \lambda_1 - n & \mbox{otherwise}.
        \end{cases}
\]
Since $\lambda_2 \leq n$, horizontal $r$-strips can always be formed for each $r$ between $r^*$ and $\lambda_1$. In conclusion, the character can be simplified further as follows:
\begin{equation} \label{char-gl-simple}
  \mathrm{char} V_\lambda = s_\lambda(x_1| \, y) = \sum_{r=r^*}^{\lambda_1} x_1^r \sum_\nu s_{\nu'}(y),
\end{equation}
where the second sum is again taken over all partitions $\nu$ such that $\lambda - \nu$ is a horizontal $r$-strip.

Like in the $\mathfrak{osp}(1|2n)$ case, a spectrum generating function can be produced. We confine ourselves to giving the energy levels of the system. They can be written as
\[
  E_k^{(\lambda)} = \hbar \omega \bigl( \frac{|\lambda|}{n-1} + r^* + k \bigr), \qquad (k=0, \ldots, \min({\lambda_1, n})).
\]
The total number of energy levels depends on $\lambda_1$ and is equal to $\min({\lambda_1, n}) + 1$.

%-------------------------------------------
% ANGULAR MOMENTUM DECOMPOSITION OF gl(1|n)
%-------------------------------------------
\section{Angular momentum decomposition of $\mathfrak{gl}(1|n)$} \label{sec-ang-gl}
We will use many of the principles of the orthosymplectic case to find generating functions for the angular momentum decomposition of $\mathfrak{gl}(1|n)$ for $n=3N$. One must always bear in mind, however, that the $\mathfrak{gl}(1|n)$ representations $V_\lambda$ are finite-dimensional. Therefore our goal will be to create a generating function in which the coefficient of $A^\lambda = A_1^{\lambda_1'} \ldots A_n^{\lambda_n'}$ represents the angular momentum decomposition of $V_\lambda$. We will thus construct a generating function comprising the angular momentum decomposition of every $\mathfrak{gl}(1|n)$ representation $V_\lambda$. This is different than our approach for $\mathfrak{osp}(1|2n)$, where the generating functions applied to just one representation $V(\mathfrak{p})$.

As before, the angular momentum operators will be part of the $\mathfrak{so}(3)$ subalgebra of $\mathfrak{gl}(1|3N)$ in the chain of subalgebras
\begin{equation} \label{so3-in-gl13N}
  \mathfrak{gl}(1|3N) \supset \mathfrak{u}(3N)
            \supset \mathfrak{u}(3) \oplus \mathfrak{u}(N) 
            \supset \mathfrak{so}(3) \oplus \mathfrak{u}(1)
\end{equation}
However, a little caution is required since it turns out that the angular momentum operators do not immediately satisfy the commutation relations of $\mathfrak{so}(3)$.

% Angular momentum
%------------------
\subsection{Angular momentum}
The angular momentum operators must obviously be defined independently from the choice of the Lie superalgebra representation. Thus, just as in equation \eqref{ang-mom-op-3}, we have
\[
  M_j = \frac{-i \hbar}{2} \sum_{k,l=1}^3 \epsilon_{jkl} \{a_k^+, a_l^- \}, \qquad (j=1,2,3)
\]
for $n=3$, i.e. the three-dimensional Wigner harmonic oscillator. In the $\mathfrak{gl}(1|3)$ solution $a_j^+ = e_{0j}$ and $a_j^- = e_{j0}$ one finds
\[
  [M_i, M_j] = \frac{i \hbar}{2} \, \epsilon_{ijk} M_k, \qquad (i,j,k = 1,2,3).
\]
So the operators $L_j = 2M_j$ generate $\mathfrak{so}(3)$. The purpose of angular momentum decomposition is mainly finding the spectrum of operators like e.g. $M^2$ or $M_3$. Obviously, this spectrum only differs by a factor from the spectrum of $L^2$ and $L_3$, so finding the $\mathfrak{so}(3)$ content is again a useful problem to tackle in this case.

For $n=3N$, the $N$-particle three-dimensional Wigner harmonic oscillator, the angular momentum operators $M_j$ are defined as in equation \eqref{ang-mom-op-3N}. Again, apart from a factor $2$ these operators generate the $\mathfrak{so}(3)$ subalgebra of $\mathfrak{gl}(1|3N)$ in the $\mathfrak{gl}(1|3N)$ solution $a_j^+ = \sqrt{2/(3N-1)} \, e_{0j}$ and $a_j^- = \sqrt{2/(3N-1)} \, e_{j0}$. Therefore we wish to know how the $\mathfrak{gl}(1|3N)$ representation $V_\lambda$ decomposes with respect to the chain of subalgebras \eqref{so3-in-gl13N}.

% Decomposing the gl(1|3N) representation V_lambda
%--------------------------------------------------
\subsection{Decomposing the $\mathfrak{gl}(1|3N)$ representation $V_\lambda$} \label{subsec-decom-gl13N}
As in the $\mathfrak{osp}(1|6N)$ case, the starting point of this decomposition is the character of $V_\lambda$, given by equation \eqref{char-gl-simple}. The Schur-functions $s_{\nu'}(y_1, \ldots, y_{3N})$ in this character are characters of the $\mathfrak{u}(3N)$ representations that occur in this decomposition. The branching to $\mathfrak{u}(3) \oplus \mathfrak{u}(N)$ representations is done by the substitution
\begin{equation}  \label{subs-gl3N-gl3glN-2}
  y_i := u_j v_l z, \qquad (j=1,2,3 \, \mbox{ and } \, l=1, \ldots, N).
\end{equation}
The $\mathfrak{u}(3)$ and $\mathfrak{u}(N)$ representations that occur in this branching can be deduced from the following known relation:
\[
    s_{\nu'}(u_1 v_1, \ldots, u_3 v_N) 
  = \sum_{\sigma, \tau} g_{\nu', \sigma, \tau} \, s_\sigma(u_1, u_2, u_3) s_\tau(v_1, \ldots, v_N),
\]
which features the Kronecker coefficients $g_{\nu', \sigma, \tau}$. These are the coefficients that appear when the product of two characters of the symmetric group $S_n$ are expanded in terms of $S_n$ characters:
\[
  \chi_\rho^{\sigma} \chi_\rho^{\tau} = \sum_{\nu'} g_{\nu', \sigma, \tau} \, \chi_\rho^{\nu'},
\]
where $\sigma, \tau, \nu'$ and $\rho$ are partitions of $n$. Recently King and Welsh~\cite{King-Welsh-11} developed a so-called ``grand generating function" for the Kronecker coefficients. Applied to our context, we can say that when $\nu', \sigma$ and $\tau$ are partitions of $n$, the Kronecker coefficients are the coefficients of the term in $z^n y^{\nu'} u^\sigma v^\tau$ in the expansion of
\begin{equation} \label{ggf-king}
  \prod_{i,j,k} \frac{1}{(1-y_i u_j v_k \, z)}
  \prod_{i<j} \left( 1-\frac{y_j}{y_i} \right)
  \prod_{i<j} \left( 1-\frac{u_j}{u_i} \right)
  \prod_{i<j} \left( 1-\frac{v_j}{v_i} \right).
\end{equation}
All products in this grand generating function run from $1$ to the length of the corresponding partitions, which in our case would be $3N$, $3$ and $N$ for $\nu'$, $\sigma$ and $\tau$ respectively. This is already a generating function for the decomposition of $\mathfrak{u}(3N)$ to $\mathfrak{u}(3) \oplus \mathfrak{u}(N)$. But the grand generating function is not a $\mathfrak{u}(3) \oplus \mathfrak{u}(N)$ representation generating function since its expansion contains many terms of the form $y^{\mu_1} u^{\mu_2} v^{\mu_3} z^n$ in which $\mu_1, \mu_2$ and $\mu_3$ are not partitions. However, the same technique as for the orthosymplectic case will turn the grand generating function into a representation generating function. This step is computationally very demanding and can only be performed for specific types of $\mathfrak{u}(3N)$ representations.

Once this step is done, the rest is easy. The $\mathfrak{u}(N)$ labels $v_1, \ldots, v_N$ need to be replaced by the dimensions of their corresponding $\mathfrak{u}(N)$ representations as before. The resulting generating function then needs to substituted into the generating function \eqref{genf-sl3-so3-pl} for the branching $\mathfrak{u}(3) \supset \mathfrak{so}(3)$. 

Generating functions for the angular momentum decomposition of the $\mathfrak{gl}(1|3N)$ representations $V_\lambda$, where $\lambda_1=1$, have been constructed for general values of $N$ in~\cite{KingSVdJ-06} using a different group theoretical method. We were able to extend these results to other forms of $\lambda$, but only for $\mathfrak{gl}(1|3)$ and $\mathfrak{gl}(1|6)$. For other values of $N$ the computations prove to be too hard.

%----------------------------------------------
% GENERATING FUNCTIONS FOR gl(1|3) AND gl(1|6)
%----------------------------------------------
\section{Generating functions for $\mathfrak{gl}(1|3)$ and $\mathfrak{gl}(1|6)$} \label{sec-gf-gl}
Since the length of the partition $\lambda$ is arbitrary, a generating function in which the coefficient of $A^\lambda = A_1^{\lambda_1} A_2^{\lambda_2} \ldots $ is the angular momentum decomposition of the $\mathfrak{gl}(1|3N)$ representation $V_\lambda$ would have an infinite amount of variables $A_i$. Therefore we choose this angular momentum decomposition to be accompanied by $A_1^{\lambda_1'} \ldots A_n^{\lambda_n'}$, thus creating a generating function with $n=3N$ variables. This is possible because only the values of $\lambda_1', \ldots, \lambda_n'$ affect the angular momentum decomposition of $V_\lambda$, as can be seen from the character formula \eqref{char-gl-simple}.

In the previous section we explained how a generating function for the decomposition of a $\mathfrak{u}(3N)$ representation in accordance with the chain of subalgebras
\[
  \mathfrak{u}(3N) \supset \mathfrak{u}(3) \oplus \mathfrak{u}(N) 
                   \supset \mathfrak{so}(3) \oplus \mathfrak{u}(1)
\]
can be created. Let us denote this generating function by $H(J, A_1, \ldots, A_n)$. The angular momentum decomposition of the $\mathfrak{gl}(1|3N)$ representation $V_\lambda$ will then be described by the following generating function:
\begin{equation} \label{factor-gl13N-u3N}
  H(J, A_1, \ldots, A_n) \; z^{\frac{|\lambda|}{n-1} + r^*} \prod_{i=1}^{\min(\lambda_1, n)} (1+A_i z).
\end{equation}
To see why this is true, we first note that the bottom energy level equals $\frac{|\lambda|}{n-1} + r^*$, which explains the power of $z$ in \eqref{factor-gl13N-u3N}. Each value of $r = r^*+k$ then defines a new energy level and is thus responsible for an extra factor $z$. For reasons of clarity, the rest of the analysis will be done for a typical partition $\lambda$, with $\lambda_1 \geq n$. The same ideas can be adopted in the atypical cases.

For $r = \lambda_1 - n$, i.e. on the ground energy level, there is only one partition $\nu$ for which $\lambda-\nu$ is a horizontal $r$-strip. Its conjugate can be written as $\nu' = (\lambda_1', \ldots, \lambda_n')$. The corresponding representation of $\mathfrak{u}(3N)$ decomposes to $\mathfrak{so}(3)$ as described by the generating function $H(J, A_1, \ldots, A_n)$, so there must be at least one term 
\[
  H(J, A_1, \ldots, A_n)
\]
in the generating function we are trying to describe. In general there are $n$ partitions $\nu$ such that $\lambda-\nu$ is a horizontal $(\lambda_1 - n+1)$-strip. Their conjugate partitions are of the form 
\[
  \nu' = (\lambda_1', \ldots, \lambda_i' - 1, \ldots, \lambda_n'),
\]
where $i=1, \ldots, n$. The angular momentum decomposition of the $\mathfrak{u}(3N)$ representations characterized by these partitions will be in the coefficient of $A_1^{\lambda_1'} \ldots A_n^{\lambda_n'}$ in $A_i \, H(J, A_1, \ldots, A_n)$. Therefore, our generating function must also contain the term
\[
  H(J, A_1, \ldots, A_n) \; z \, (A_1 + \cdots + A_n).
\]
Note that it is possible that $\lambda_i' = \lambda_{i+1}'$, in which case $\nu'$ would not be a partition. However, in this case the coefficient of $A^{\nu'}$ in $H(J, A_1, \ldots, A_n)$ will be zero. In other words, this ``non-partition'' will not be counted at all.

The factor $(A_1 + \cdots + A_n)$ in the term for $r=\lambda_1 - n+1$ is in fact the elementary symmetric polynomial $e_1(A_1, \ldots, A_n)$. The elementary symmetric polynomial $e_i$ in $n$ variables is defined by the sum of all possible products of $i$ out of the $n$ variables. We refer to Macdonald~\cite{MacDonald} for a more elaborate introduction to elementary symmetric polynomials. For $r = \lambda_1 - n+2$ there are typically $n(n-1)/2$ partitions $\nu$ for which $\lambda-\nu$ is a horizontal $r$-strip. An analogous reasoning as before shows that they are responsible for a term
\[
  H(J, A_1, \ldots, A_n) \; z^2 \, \sum_{1 \leq i<j \leq n} A_i A_j
\]
in our generating function. This term contains the elementary symmetric function $e_2(A_1, \ldots, A_n)$. Taking all values of $r$ into account, it is not so hard to see that the generating function for the angular momentum decomposition of the $\mathfrak{gl}(1|3N)$ representation $V_\lambda$ equals
\[
  H(J, A_1, \ldots, A_n) \; z^{\frac{|\lambda|}{n-1} + \lambda_1 - n} \sum_{i=0}^n e_i(A_1, \ldots, A_n) z^i
\]
in the typical case where $\lambda_1 \geq n$. In this expression we find back the generating function for elementary symmetric functions, which can be rewritten (see for example Macdonald~\cite{MacDonald}) as
\[
  \sum_{i=0}^n e_i(A_1, \ldots, A_n) z^i = \prod_{i=1}^n (1 + A_i z).
\]
For the atypical cases, where $\lambda_1 < n$, we can build up a similar analysis to eventually obtain the generating function in equation \eqref{factor-gl13N-u3N}.

Clearly, the most important part of our problem is finding the $\mathfrak{u}(3N) \supset \mathfrak{so}(3)$ generating function $H(J, A_1, \ldots, A_n)$. However, for $\mathfrak{gl}(1|3)$ this step is trivial, so the results can be written down immediately.

% Generating functions for gl(1|3)
%----------------------------------
\subsection{Generating functions for $\mathfrak{gl}(1|3) \supset \mathfrak{so}(3)$}
For the main part, the decomposition of the representation $V_\lambda$ of $\mathfrak{gl}(1|3)$ following the branching
\[
  \mathfrak{gl}(1|3) \supset \mathfrak{u}(3) \supset \mathfrak{so}(3)
\]
is described by the generating function $G(A_1, A_2, A_3)$ given by equation \eqref{genf-sl3-so3-pl}, where $A_1$, $A_2$ and $A_3$ label the first three parts of the conjugate partition of $\lambda$. The rest of the generating function follows from the previous discussion, equation \eqref{factor-gl13N-u3N} in particular, and depends on $\lambda_1$. For $\lambda_1 \geq 3$ we find
\begin{equation} \label{gf-ang-gl13}
  z^{\frac{|\lambda|}{2} + r^*} 
  \frac{(1+A_1 z)(1+A_2 z)(1+A_3z)(1+A_1^2 A_2 \, J)}
       {(1 - A_1 A_2 A_3)(1 - A_1 \, J)(1 - A_1 A_2 \, J)(1 - A_1^2)(1 - A_1^2 A_2^2)}.
\end{equation}
The cases where $\ell(\lambda')=2$ and $\ell(\lambda')=1$ are easily deduced from this equation by setting $A_3=0$ and $A_2=A_3=0$ respectively. For $\ell(\lambda')=1$ we find back the results from King, Stoilova and Van der Jeugt in~\cite{KingSVdJ-06}.

The generating function \eqref{gf-ang-gl13} allows us to construct $(E,j)$-diagrams for any $\mathfrak{gl}(1|3)$ representation $V_\lambda$. Some examples for the typical case are given below, with $\lambda_a = (3,1,0)$ and  $\lambda_b = (3,2,2,1,1)$.
\[
  \begin{tabular}{cc|ccc}
             5        &     &   & 1 &   \\
             4        &     & 1 & 1 & 1 \\
             3        &     & 1 & 1 & 1 \\
             2        &     &   & 1 &   \\ \hline 
    $E_k^{\lambda_a}$ &     &   &   &   \\
                      & $j$ & 0 & 1 & 2  
  \end{tabular}
%  %
%  \qquad
%  %
%  \begin{tabular}{cc|ccccc}
%             7        &     &   & 1 & 1 & 1 &   \\
%             6        &     & 1 & 2 & 3 & 2 & 1 \\
%             5        &     & 1 & 2 & 3 & 2 & 1 \\
%             4        &     &   & 1 & 1 & 1 &   \\ \hline 
%    $E_k^{(3,2,2,1)}$ &     &   &   &   &   &   \\
%                      & $j$ & 0 & 1 & 2 & 3 & 4  
%  \end{tabular}
  %
  \qquad \mbox{ and } \qquad
  \begin{tabular}{cc|cccccc}
          15/2        &     & 1 &   & 2 & 1 & 1 &   \\
          13/2        &     &   & 3 & 3 & 4 & 2 & 1 \\
          11/2        &     &   & 3 & 3 & 4 & 2 & 1 \\
           9/2        &     & 1 &   & 2 & 1 & 1 &   \\ \hline 
    $E_k^{\lambda_b}$ &     &   &   &   &   &   &   \\
                      & $j$ & 0 & 1 & 2 & 3 & 4 & 5   
  \end{tabular}
\]
The information of these $(E,j)$-diagrams can be obtained from the coefficient of $A^{\nu_a'}$ and $A^{\nu_b'}$ in the expansion of \eqref{gf-ang-gl13}, for $\nu_a' = (2,1,1)$ and $\nu_b' = (5,3,1)$ respectively. As a primary difference with the canonical case (and the $\mathfrak{osp}(1|6N)$ case in general) we see that there is a finite amount of energy levels. Also, at the bottom energy level we see more than one $\mathfrak{so}(3)$-multiplet. The first $(E,j)$-diagram, where $\lambda = (3,1,0)$ shows an exception to this remark. We also note that we still have equidistant energy levels, and that there are again higher multiplicities of $\mathfrak{so}(3)$ representations.

In the atypical cases, the number of energy levels decreases as the length of $\lambda$ becomes smaller. For $\lambda = (2,1,1)$ we have
\[
  \begin{tabular}{cc|cccc}
             4        &     & 1 &   & 1 &    \\
             3        &     &   & 2 & 1 & 1  \\
             2        &     &   & 1 & 1 & 1  \\ \hline 
    $E_k^{\lambda}$   &     &   &   &   &    \\
                      & $j$ & 0 & 1 & 2 & 3  
  \end{tabular}
\]
There are only three energy levels in this case, and we observe that the vertical symmetry of the $(E,j)$-diagram is now gone.

% Generating functions for gl(1|6)
%----------------------------------
\subsection{Generating functions for $\mathfrak{gl}(1|6) \supset \mathfrak{so}(3)$}
The case $\mathfrak{gl}(1|3)$ is deceivingly simple compared to $\mathfrak{gl}(1|6)$. In fact, it will no longer be possible to construct the generating function $H(J, A_1, \ldots, A_6)$ for all representations of $\mathfrak{u}(6)$. More precisely, we will only be able to handle the cases $\ell(\lambda')=1$ and $\ell(\lambda')=2$ completely, where $\lambda$ is the partition that characterizes the $\mathfrak{gl}(1|6)$ representation $V_\lambda$. 

% The representation V_lambda with l(lambda')=1
%- - - - - - - - - - - - - - - - - - - - - - - -
\subsubsection*{The representation $V_\lambda$ with $\ell(\lambda')=1$}
We follow the procedure described in section \ref{subsec-decom-gl13N} to obtain a generating function $H(J, A_1)$ which will describe the angular momentum generating function for the $\mathfrak{u}(6)$ representation characterized by a partition $\nu'$, where $\nu'$ contains of one part only. The starting point is the grand generating function \eqref{ggf-king} for $N=2$:
\[
  \frac{ (1-\frac{u_2}{u_1})(1-\frac{u_3}{u_1})(1-\frac{u_3}{u_2}) (1-\frac{v_2}{v_1}) }
       { (1-u_1 v_1 A_1)(1-u_2 v_1 A_1)(1-u_3 v_1 A_1)(1-u_1 v_2 A_1)(1-u_2 v_2 A_1)(1-u_3 v_2 A_1) },
\]
where we have taken into account the fact that $\lambda_1 = \ell(\lambda') = 1$ and therefore only one parameter $A_1$ is necessary to describe the $\mathfrak{u}(6)$ representation. We recognize this function from the $V(1)$ representation of $\mathfrak{osp}(1|12)$, where the starting function was the same. Thus, the rest of the analysis can be adopted from that case and eventually we find
\[
  H(J, A_1) = \frac{(1+JA_1^2)}{(1-A_1^2)^3 (1-JA_1)^2}.
\]
The generating function for the angular momentum decomposition of the $\mathfrak{gl}(1|6)$ representation $V_\lambda$, with $\ell(\lambda')=1$, then follows from equation \eqref{factor-gl13N-u3N} and is equal to
\begin{equation} \label{GF-gl16-lambda1-1}
  z^\frac{|\lambda|}{5} \, \frac{(1+A_1z) \, (1+JA_1^2)}{(1-A_1^2)^3 \, (1-JA_1)^2}.
\end{equation}
This is again confirmed by the results in~\cite{KingSVdJ-06}.

% The representation V_lambda with l(lambda')=2
%- - - - - - - - - - - - - - - - - - - - - - - -
\subsubsection*{The representation $V_\lambda$ with $\ell(\lambda')=2$}
Compared to the case where $\ell(\lambda')=1$, the denominator of the grand generating function will have six extra factors containing $A_2$, and the numerator must have an extra factor $(1-A_2/A_1)$. From that point on, essentially all computations run along previously traveled paths. Yet, the computer has a much harder time performing these computations, and we have the end result (see Appendix~A) as a witness. Some intermediate results are however interesting. The $\mathfrak{u}(3) \oplus \mathfrak{u}(2)$ representation generating function in this case equals $N/D$ with 
\begin{subequations} \label{u3-u2-rep-gf-Vlambda2}
  \begin{equation}
    \begin{split}
      N = \, 1 & + \, u^{21}v^{21}A^{21} + u^{221}v^{32}A^{32} + u^{321}v^{42}A^{33} 
                                         + u^{321}v^{33}A^{42} - 2u^{321}v^{42}A^{42}  \\
               & - \, u^{421}v^{52}A^{43} - u^{421}v^{43}A^{52} - u^{422}v^{53}A^{53} 
                                          - u^{431}v^{53}A^{53} - u^{432}v^{63}A^{54}  \\
               & - \, u^{432}v^{54}A^{63} - 2u^{532}v^{64}A^{64} + u^{532}v^{73}A^{64} 
                                          + u^{532}v^{64}A^{73} + u^{632}v^{74}A^{74}  \\
               & + \, u^{643}v^{85}A^{85} + u^{853}v^{10,6}A^{10,6}
    \end{split}
  \end{equation}
  and
  \begin{equation}
    \begin{split}
      D = \;\, & (1 - u^{1}v^{1}A^{1}) (1 - u^{2}v^{11}A^{11}) (1 - u^{11}v^{2}A^{11})
                                       (1 - u^{11}v^{11}A^{2}) (1 - u^{111}v^{21}A^{21})                                 \\
               & (1 - u^{22}v^{22}A^{22}) (1 - u^{211}v^{31}A^{22}) (1 - u^{211}v^{22}A^{31}) (1 - u^{222}v^{33}A^{33}),
    \end{split}
  \end{equation}
\end{subequations}
where we have used the notation $u^\sigma = u_1^{\sigma_1} u_2^{\sigma_2} u_3^{\sigma_3}$, and similarly for $v$ and $A$. This generating function was first obtained by Patera and Sharp~\cite{Patera-Sharp-80} as a plethysm generating function for two-rowed representations of $SU(n)$. In this paper, we have already computed $N/D$ given by equation \eqref{u3-u2-rep-gf-Vlambda2} independently, not using the grand generating function of King and Welsh. Indeed, this generating function and the function $N/D$, with numerator and denominator given by equation \eqref{u3-u2-rep-gf-V2}, are similar. The only difference is that $A^\lambda$ in \eqref{u3-u2-rep-gf-Vlambda2} has been changed into $z^{|\lambda|}$ in equation \eqref{u3-u2-rep-gf-V2}. The functions $N/D$ given by equations \eqref{u3-u2-rep-gf-V2} and \eqref{u3-u2-rep-gf-Vlambda2} represent the $\mathfrak{u}(3) \oplus \mathfrak{u}(2)$ branching of all $\mathfrak{u}(6)$ representations occurring in the representation $V(2)$ of $\mathfrak{osp}(1|12)$ and $V_\lambda$ (with $\ell(\lambda')=2$) of $\mathfrak{gl}(1|6)$ respectively. The $\mathfrak{u}(6)$ representations that occur in both cases are characterized by a partition with a maximum length of two, so both generating functions must be equal.

Introducing the dimensions of the $\mathfrak{u}(2)$ representations in our $\mathfrak{u}(3) \oplus \mathfrak{u}(2)$ representation generating function, and then substituting the result into $G(u_1, u_2, u_3)$ gives us the function $H(J, A_1, A_2)$. The unappealing sight of this function forces us to relocate its full expression to Appendix~A. We can still write our angular momentum content generating function as
\[
  z^\frac{|\lambda|}{5} \, (1+A_1z)(1+A_2z) \, H(J, A_1, A_2).
\]
Note that for $A_2=0$ we must find back equation \eqref{GF-gl16-lambda1-1}, which is indeed the case. It is interesting to see what happens when the powers of $A_1$ and $A_2$ are equal. This means that we are looking at $\mathfrak{u}(6)$ representations with character $s_{\nu'}(A_1, A_2)$, where $\nu'$ is a partition for which both rows have equal length. These representations are of interest in complexity theory and in the study of qubits~\cite{Luque-Thybon-2003, Luque-Thybon-2006}. The generating function in this case can be computed out of the previous one by making the substitution
\[
  A_1 = aA_1, \qquad A_2 = \frac{A_2}{a}
\]
and then looking for the constant term in $a$. We find that the $\mathfrak{u}(6) \supset \mathfrak{u}(3) \oplus \mathfrak{u}(2)$ branching is represented by the generating function
\[
  \frac{ (1 + u^{321}v^{42}A^{33}) }
       { (1 - u^{2}v^{11}A^{11}) (1 - u^{11}v^{2}A^{11}) (1 - u^{22}v^{22}A^{22})
                                 (1 - u^{211}v^{31}A^{22}) (1 - u^{222}v^{33}A^{33}) },
\]
a result which was obtained recently by King and Welsh~\cite{King-Welsh-11}. This confirms earlier observations concerning inner products of Schur functions~\cite{Garsia-2009, Brown-Willigenburg-Zabrocki-2010}. The generating function $H_{\lambda_1' = \lambda_2'}(J, A_1, A_2) = N/D$, with
\beqas
  N = 1 & - \, & A + 2A^{22} + 3JA^{22} + 3J^2A^{22} - A^{33} - 3JA^{33} - 4J^2A^{33} - 3J^3A^{33} + A^{44}      \\
        & + \, & 3JA^{44} - 2J^2A^{44} + 3J^3A^{44} + J^4A^{44} - 3JA^{55} - 4J^2A^{55} - 3J^3A^{55} - J^4A^{55} \\
        & + \, & 3J^2A^{66} + 3J^3A^{66} + 2J^4A^{66} - J^4A^{77} + J^4A^{88}
\eeqas
and
\[
  D = (1-A^{11})^2 \, (1-JA^{11})^3 \, (1-J^2A^{11}) \, (1-A^{22})^4,
\]
describes the further branching of a two-rowed $\mathfrak{u}(6)$ representation to $\mathfrak{so}(3)$. Note that we cannot use this generating function to describe the angular momentum decomposition of the $\mathfrak{gl}(1|6)$ representation $V_\lambda$, for which $\lambda$ is a partition with two columns of equal length in the Young diagram. Indeed, solving such a problem would require the angular momentum decomposition of all $\mathfrak{u}(6)$ representations characterized by a partition $\nu'$, such that $\lambda - \nu$ is a horizontal $r$-strip, with $r=0, 1, 2$. For $r=1$, the partition $\nu'$ does not consist of two rows of equal length (in fact, $\nu_1' = \nu_2' + 1$), thus the generating function $H_{\lambda_1' = \lambda_2'}(J, A_1, A_2)$ is of no use for $\mathfrak{u}(6)$ representations characterized by this particular partition $\nu'$.

% The representation V_lambda with l(lambda')=3
%- - - - - - - - - - - - - - - - - - - - - - - -
\subsubsection*{The representation $V_\lambda$ with $\ell(\lambda')=3$}
Computationally, this case can only be worked out when $\lambda_1' = \lambda_2' = \lambda_3'$. A three-rowed $\mathfrak{u}(6)$ representation where all rows are of equal length decomposes to $\mathfrak{u}(3) \oplus \mathfrak{u}(2)$ in accordance with the generating function of the form $N/D$, with
\beqas
  N = 1 & - \, & u^{111}v^{21}A^{111} - u^{321}v^{33}A^{222} + u^{222}v^{42}A^{222} 
                                      + u^{321}v^{42}A^{222} + 2u^{432}v^{54}A^{333}           \\
        & + \, & u^{531}v^{54}A^{333} - u^{432}v^{63}A^{333} - v^{75}u^{543}A^{444} 
                                      + u^{642}v^{66}A^{444} - 2u^{642}v^{75}A^{444}           \\
        & - \, & u^{753}v^{87}A^{555} - u^{852}v^{87}A^{555} + u^{753}v^{96}A^{555} 
                                      + u^{963}v^{10, 8}A^{666} - u^{10, 7, 4}v^{12, 9}A^{777}
\eeqas
and
\beqas
  D & = & (1-u^{111}v^{21}A^{111}) (1-u^{21}v^{21}A^{111}) (1-u^{111}v^{3}A^{111}) (1-u^{321}v^{33}A^{222}) \\
    &   & \times \, (1-u^{33}v^{33}A^{222}) (1-u^{411}v^{33}A^{222}) (1-u^{444}v^{66}A^{444})
\eeqas
The same representation of $\mathfrak{u}(6)$ decomposes to $\mathfrak{so}(3)$ in accordance with the generating function $H_{\lambda_1' = \lambda_2' = \lambda_3'}(J, A_1, A_2, A_3) = N/D$, with
\beqas
  N = 1 & + \, & 2JA^{222} + 3J^2A^{222} + J^3A^{222} + 2JA^{333} - 2J^2A^{333} - 2J^3A^{333} - 2J^4A^{333} \\
        & + \, & A^{444} - 4J^2A^{444} - 4J^3A^{444} + J^5A^{444} - 2JA^{555} - 2J^2A^{555} - 2J^3A^{555}   \\
        & + \, & 2J^4A^{555} + J^2A^{666} + 3J^3A^{666} + 2J^4A^{666} + J^5A^{888}
\eeqas
and
\[
  D = (1-A^{111})^4 \, (1-JA^{111})^2 \, (1-J^2A^{111})^2 \, (1-A^{222})^3.
\]
The argumentation given in the previous section implies again that this generating function does not contain sufficient information for the angular momentum decomposition of the $\mathfrak{gl}(1|6)$ representation $V_\lambda$, with $\lambda_1' = \lambda_2' = \lambda_3'$.

%-------------
% CONCLUSIONS
%-------------
\section{Conclusions}
For a 3D $N$-particle Wigner harmonic oscillator, operator solutions exist in terms of generators of the Lie superalgebras $\mathfrak{osp}(1|6N)$ and $\mathfrak{gl}(1|3N)$. These operators act in representation spaces of these Lie superalgebras. Our goal was to find the angular momentum/energy contents of the representations $V(\mathfrak{p})$ of $\mathfrak{osp}(1|6N)$ and $V_\lambda$ of $\mathfrak{gl}(1|3N)$. For $N=1$, we have been able to construct generating functions representing the angular momentum decomposition for all of these representations. For $N=2$, the computer allowed us to construct only partial results. For $\mathfrak{osp}(1|12)$ we have generating functions for the representations $V(1)$ and $V(2)$, but for other representations the results proved computationally too hard. In the $\mathfrak{gl}(1|6)$ case, we had to restrict ourselves to representations $V_\lambda$ for which $\lambda$ had a maximum of two columns.

By means of the obtained generating functions, we were able to plot the angular momentum/energy contents in so-called $(E,j)$-diagrams. These are tables showing the multiplicities of all angular momentum values at each energy level. These tables are a practical tool to compare the results of the new Wigner solutions to the well-known canonical case.

For the 1-dimensional Wigner harmonic oscillator, investigated by Wigner in~\cite{Wigner}, the energy levels in the non-canonical solutions are shifted in height but remain equidistant. The $(E,j)$-diagrams in the $\mathfrak{osp}(1|6N)$ solution show that for all representations $V(\mathfrak{p})$ we have a similar behavior for the angular momentum contents. Apart from the shifted energy levels, the structure of the $(E,j)$-diagrams in the non-canonical solutions is the same as that for the representation $V(1)$ of $\mathfrak{osp}(1|6)$. The main difference is that the multiplicities of the angular momentum representations can be higher than 1, a feature not observed in the canonical case.

In the $\mathfrak{gl}(1|3N)$ solution, the situation is drastically different due to the finite-dimensional nature of the representations. There is a finite amount of energy levels and the number of $\mathfrak{so}(3)$-multiplets, the angular momentum contents, does not increase when the energy gets higher. On the contrary, for higher (and lower) energy levels we see less $\mathfrak{so}(3)$-multiplets than in the bulk of the energy spectrum.

%-----------------
% ACKNOWLEDGMENTS
%-----------------
\subsection*{Acknowledgments}
We wish to thank Prof. R. C. King for a fruitful and enlightening visit to Ghent University.
G.\ Regniers was supported by project P6/02 of the Interuniversity Attraction Poles Programme (Belgian State -- 
Belgian Science Policy).

%\appendix

% THE FUNCTION H(J, A1, A2)
%---------------------------
\section*{Appendix A. The function $H(J, A_1, A_2)$} %\label{app-HJA1A2}
The generating function $H(J, A_1, A_2)$ for the angular momentum decomposition of a two-rowed representation of $\mathfrak{u}(6)$, has the form $N/D$, with
\beqas
N = \, 1 & - \, & A^{11} + JA^2 + 2A^{21} + 6JA^{21} + 2A^{22} + 3JA^{22} + 3J^2A^{22} + 3A^{31} - 2JA^{31}            \\
         & - \, & 6J^2A^{31} - J^3A^{31} + 2A^{32} - 6JA^{32} - 8J^2A^{32} - 6J^3A^{32} - A^{33} - 3JA^{33}            \\
         & - \, & 4J^2A^{33} - 3J^3A^{33} - 6JA^{41} - 3A^{42} - 6JA^{42} - 6J^2A^{42} + 5J^3A^{42} + 3J^4A^{42}       \\
         & - \, & 2JA^{43} + 4J^2A^{43} + 8J^3A^{43} + 6J^4A^{43} + A^{44} + 3JA^{44} - 2J^2A^{44} + 3J^3A^{44}        \\
         & + \, & J^4A^{44} - A^{51} + 3J^2A^{51} + 4JA^{52} + 6J^3A^{52} + 3A^{53} + 2JA^{53} + J^2A^{53}             \\
         & + \, & 3J^3A^{53} - 4J^4A^{53} - 3J^5A^{53} - 2JA^{54} - 8J^2A^{54} - 6J^4A^{54} - 2J^5A^{54}               \\
         & - \, & 3JA^{55} - 4J^2A^{55} - 3J^3A^{55} - J^4A^{55} + 2JA^{61} + A^{62} + J^2A^{62} - 2A^{63}             \\
         & - \, & 10JA^{63} - 10J^2A^{63} - 3A^{64} - 11JA^{64} + 4J^2A^{64} + 5J^3A^{64} + 6J^4A^{64} + 3J^5A^{64}    \\
         & + \, & J^6A^{64} + 4J^2A^{65} + 6J^3A^{65} + 8J^4A^{65} + 2J^5A^{65} + 3J^2A^{66} + 3J^3A^{66}              \\
         & + \, & 2J^4A^{66} - J^2A^{71} - 2JA^{72} - 2J^3A^{72} - 2A^{73} + JA^{73} + 8J^2A^{73} + 8J^3A^{73}         \\
         & + \, & 6J^4A^{73} - 2A^{74} + 8JA^{74} + 30J^2A^{74} + 18J^3A^{74} + 8J^4A^{74} - 4J^5A^{74} + 8JA^{75}     \\
         & + \, & 15J^2A^{75} + 12J^3A^{75} + 3J^4A^{75} - 7J^5A^{75} - J^6A^{75} - 2J^3A^{76} - 4J^4A^{76}            \\
         & - \, & 4J^5A^{76} - J^4A^{77} + J^2A^{82} + 4JA^{83} + 4J^2A^{83} + 2J^3A^{83} + A^{84} + 7JA^{84}          \\
         & - \, & 3J^2A^{84} - 12J^3A^{84} - 15J^4A^{84} - 8J^5A^{84} + 4JA^{85} - 8J^2A^{85} - 18J^3A^{85}            \\
         & - \, & 30J^4A^{85} - 8J^5A^{85} + 2J^6A^{85} - 6J^2A^{86} - 8J^3A^{86} - 8J^4A^{86} - J^5A^{86}             \\
         & + \, & 2J^6A^{86} + 2J^3A^{87} + 2J^5A^{87} + J^4A^{88} - 2J^2A^{93} - 3J^3A^{93} - 3J^4A^{93}              \\
         & - \, & 2JA^{94} - 8J^2A^{94} - 6J^3A^{94} - 4J^4A^{94} - A^{95} - 3JA^{95} - 6J^2A^{95} - 5J^3A^{95}        \\
         & - \, & 4J^4A^{95} + 11J^5A^{95} + 3J^6A^{95} + 10J^4A^{96} + 10J^5A^{96} + 2J^6A^{96} - J^4A^{97}           \\
         & - \, & J^6A^{97} - 2J^5A^{98} + J^2A^{10,4} + 3J^3A^{10,4} + 4J^4A^{10,4} + 3J^5A^{10,4} + 2JA^{10,5}       \\
         & + \, & 6J^2A^{10,5} + 8J^4A^{10,5} + 2J^5A^{10,5} + 3JA^{10,6} + 4J^2A^{10,6}  - 3J^3A^{10,6} - J^4A^{10,6} \\
         & - \, & 2J^5A^{10,6} - 3J^6A^{10,6} - 6J^3A^{10,7} - 4J^5A^{10,7} - 3J^4A^{10,8} + J^6A^{10,8} - J^2A^{11,5} \\
         & - \, & 3J^3A^{11,5} + 2J^4A^{11,5} - 3J^5A^{11,5} - J^6A^{11,5} - 6J^2A^{11,6} - 8J^3A^{11,6}               \\
         & - \, & 4J^4A^{11,6} + 2J^5A^{11,6} - 3J^2A^{11,7} - 5J^3A^{11,7} + 6J^4A^{11,7} + 6J^5A^{11,7}              \\
         & + \, & 3J^6A^{11,7} + 6J^5A^{11,8} + 3J^3A^{12,6} + 4J^4A^{12,6} + 3J^5A^{12,6} + J^6A^{12,6}               \\
         & + \, & 6J^3A^{12,7} + 8J^4A^{12,7} + 6J^5A^{12,7} - 2J^6A^{12,7} + J^3A^{12,8} + 6J^4A^{12,8}               \\ 
         & + \, & 2J^5A^{12,8} - 3J^6A^{12,8} - 3J^4A^{13,7} - 3J^5A^{13,7} - 2J^6A^{13,7} - 6J^5A^{13,8}              \\
         & - \, & 2J^6A^{13,8} - J^5A^{13,9} + J^6A^{14,8} - J^6A^{15,9}
\eeqas
and
\[
  D = (1-A^{11})^2 (1-A^2)^3 (1-A^{22})^4 (1-JA)^2 (1-JA^{11})^3 (1-J^2A^{11}).
\]
We have used the notation $A^{\nu'}$ to denote $A_1^{\nu_1'} A_2^{\nu_2'}$ for practical reasons.

\newpage
%------------
% REFERENCES
%------------


\begin{thebibliography}{99}

\bibitem{Wigner}
E. P.\ Wigner, 
% Do the equations of motion determine the quantum mechanical commutation relations? 
Phys.\ Rev. {\bf 77}, 711-712 (1950).

\bibitem{Palev-79}
T.D.\ Palev,
% Lie-superalgebraical approach to the second quantization
Czech J.\ Phys., Sect. {\bf B29}, 91-98 (1979).

\bibitem{Palev-82}
T.D.\ Palev,
% Wigner approach to quantization. Noncanonical quantization of two particles interacting via a harmonic potential 
J.\ Math.\ Phys. {\bf 23}, 1778-1784 (1982).

\bibitem{Palev-86}
A.H.\ Kamupingene, T.D.\ Palev and S.P.\ Tsavena,
% Wigner quantum systems. Two particles interacting via a harmonic potential. I. Two-dimensional space 
J.\ Math.\ Phys. {\bf 27}, 2067-2075 (1986).

\bibitem{LVdJ-08}
S.\ Lievens and J.\ Van der Jeugt,
% Spectrum generating functions for non-canonical quantum oscillators
J.\ Phys.\ A: Math.\ Theor. {\bf 41}, 355204 (2008).

\bibitem{Ganchev}
Ganchev A.Ch., Palev T.D.,
%A Lie superalgebraic interpretation of the para-Bose statistics,
{\it J.\ Math.\ Phys.} {\bf 21} (1980), 797--799.

\bibitem{LSVdJ-08-2}
S.\ Lievens, N.I.\ Stoilova and J.\ Van der Jeugt,
% The paraboson Fock space and unitary irreducible representations of the Lie superalgebra osp(1|2n) 
Comm.\ Math.\ Phys. {\bf 281}, 805-826 (2008).

\bibitem{MacDonald}
I.\ G. Macdonald,
Symmetric functions and Hall polynomials,
Oxford University Press, Oxford, 2nd edition (1995).

\bibitem{Littlewood-40}
D.\ E. Littlewood,
The theory of group characters,
Oxford University Press, Oxford, (1940).

\bibitem{Gaskell-78}
R. Gaskell, A. Peccia and R.T. Sharp,
% Generating functions for polynomial irreducible tensors
J.\ Math.\ Phys. {\bf 19}, 727 (1978).

\bibitem{Wybourne-74}
B.G. Wybourne,
Classical groups for physicists
Wiley, New York (1978).

\bibitem{Gould}
M. D.\ Gould and R. B.\ Zhang, 
% Classification of all star irreps of gl(m|n)
J.\ Math.\ Phys. {\bf 31}, 2552-2559 (1990).

\bibitem{Berele-87}
A. Berele and A. Regev,
% Hook Young-diagrams with applications to combinatorics and to representations of Lie-superalgebras,
Adv. Math. {\bf 64}, 118--175, (1987).

\bibitem{King-Welsh-11}
R.\ C. King and T.\ A. Welsh,
Symmetric group characters and generating functions for Kronecker and reduced Kronecker coefficients,
private communication

\bibitem{KingSVdJ-06}
R.\ C. King, N.\ I.\ Stoilova and J.\ {Van der Jeugt},
% Representations of the Lie superalgebra $\mathfrak{gl}(1|n)$ in a Gelfand-Zetlin basis and Wigner quantum oscillators 
J.\ Phys.\ A: Math.\ Gen. {\bf 39}, 5763-5785 (2006).

\bibitem{Patera-Sharp-80}
J.\ Patera and R.\ T. Sharp,
% Generating functions for plethysms of finite and continuous groups 
J.\ Phys.\ A: Math.\ Gen. {\bf 13}, 397-416 (1980).

\bibitem{Luque-Thybon-2003}
J.-G.\ Luque and J.-Y.\ Thybon,
% Polynomial invariants of four qubits
Phys. Rev. A {\bf 67}, 042303 (2003).

\bibitem{Luque-Thybon-2006}
J.-G.\ Luque and J.-Y.\ Thybon,
% Algebraic invariants of five qubits 
J.\ Phys.\ A: Math.\ Gen {\bf 39}, 371-377 (2006).

\bibitem{Garsia-2009}
A.\ Garsia, N.\ Wallach, G.\ Xin and M.\ Zabrocki,
%  Kronecker coefficients via symmetric functions and constant term identities, 
to appear.

\bibitem{Brown-Willigenburg-Zabrocki-2010}
A.\ Brown, S.\ van Willigenburg and M.\ Zabrocki,
% Expressions for Catalan Kronecker products 
Pacific J. Math. {\bf 248}, 31-48 (2010).

\end{thebibliography}
\end{document}